\documentclass[notitlepage,a4paper,aps,prd,onecolumn,superscriptaddress,nofootinbib,groupedaddress]{revtex4}
\usepackage{amsmath}
\usepackage{amsfonts,color}
\usepackage{amsmath}
\usepackage{amsthm}
\usepackage{pdfpages}
\usepackage{amsfonts,color}
\usepackage{amssymb,float}
\usepackage{verbatim}
\usepackage{booktabs,siunitx}

\usepackage{amsmath}

\newcommand{\dd}{{\rm d}}
\usepackage[utf8]{inputenc}
\allowdisplaybreaks
\usepackage{color}
\usepackage{hyperref}
\hypersetup{
    colorlinks=true,
    linkcolor=blue,
    filecolor=magenta,      
   citecolor=blue
}

\usepackage{enumitem}

\usepackage{tikz}
\usetikzlibrary{shapes.geometric}
\usetikzlibrary{arrows.meta,arrows}

\begin{document}
\title{Cosmology of Cubic Poincaré Gauge gravity}

\author{Sebastian Bahamonde}
\email{sbahamondebeltran@gmail.com}
\affiliation{Kavli Institute for the Physics and Mathematics of the Universe (WPI), The University of Tokyo Institutes
for Advanced Study (UTIAS), The University of Tokyo, Kashiwa, Chiba 277-8583, Japan.}
\affiliation{Cosmology, Gravity, and Astroparticle Physics Group, Center for Theoretical Physics of the Universe,
Institute for Basic Science (IBS), Daejeon, 34126, Korea.}

\author{Rebecca Briffa}
\email{rebecca.briffa.16@um.edu.mt}
\affiliation{Institute of Space Sciences and Astronomy, University of Malta, Msida, Malta and
Department of Physics, University of Malta, Msida, Malta}

\author{Konstantinos Dialektopoulos}
\email{kdialekt@gmail.com}
\affiliation{Department of Mathematics and Computer Science, Transilvania University of Brasov, Eroilor 29, Brasov, Romania}
\affiliation{Institute of Space Sciences and Astronomy, University of Malta, Msida, Malta and Department of Physics, University of Malta, Msida, Malta}

\author{Damianos Iosifidis}
\email{d.iosifidis@ssmeridionale.it}

   \affiliation{Scuola Superiore Meridionale, Largo San Marcellino 10, 80138 Napoli, Italy}
    
\affiliation{INFN– Sezione di Napoli, Via Cintia, 80126 Napoli, Italy}

\author{Jackson Levi Said}
\email{jackson.said@um.edu.mt}
\affiliation{Institute of Space Sciences and Astronomy, University of Malta, Msida, Malta and
Department of Physics, University of Malta, Msida, Malta}

\begin{abstract}
In this paper, we study flat FLRW cosmology for a Poincaré gauge theory containing cubic invariants that is free from ghosts in arbitrary backgrounds in the axial and vector sectors of the torsion tensor. The new degrees of freedom can be related to hypermomentum but continue to be dynamical even in vacuum. These extra degrees of freedom open a more natural way in which to construct potential gravitational models that provide possible ways to modify astrophysical and cosmological physics. In this framework, we study two particular branches of the theory where preliminary routes of exploring these new variables are exposed. The first is the branch where the hypermomentum vanishes, while the second branch involves the setting where the  perfect fluid  and hypermomentum parts of the sources are independently conserved. In both settings, we find generically faster expanding cosmologies with similar estimates of the cosmic matter content as in the standard model of cosmology. Cubic Poincaré Gauge gravity offers an interesting theoretical basis on which to study cosmology, and indicates some preliminary positive constraints when compared with observational constraints.
\end{abstract}

\maketitle

\section{Introduction}
The $\Lambda$CDM model has endured over two decades as the cosmological concordance model with extreme agreement with many observational measurements that span all cosmic scales \cite{Misner:1974qy,Clifton:2011jh}. Here, cold dark matter (CDM) acts as a stabilizing both for perturbations in the early Universe \cite{Peebles:2002gy} and for galactic structures in the late Universe \cite{Baudis:2016qwx,Bertone:2004pz}. On the other hand, dark energy is realized through the cosmological constant \cite{Copeland:2006wr} and dominates in the late accelerating Universe. Despite these efforts, internal consistency issues with the cosmological constant \cite{Weinberg:1988cp} have still not been resolved, and viable particle physics descriptions remain elusive in detectors \cite{Gaitskell:2004gd}. Moreover, anomalies in the $\Lambda$CDM evolution of the Universe continue to remain open such as the missing Lithium \cite{DiBari:2013dna}, large scale anomalies in the cosmic microwave background radiation (CMB) ranging from the lack of large-angle CMB temperature correlations \cite{doi:10.1073/pnas.90.11.4766} to the measured asymmetries \cite{2020ApJS..247...69E,2007ApJ...660L..81E}, and others \cite{Aghanim:2018eyx,Perivolaropoulos:2021jda}. Recently, an even more severe problem has arisen calling into question the effectiveness of the concordance model, namely cosmic tensions \cite{CosmoVerse:2025txj,DiValentino:2020vhf,DiValentino:2020zio,DiValentino:2020vvd,DiValentino:2020srs}. The statistical tension is between direct measurements of the Hubble tension which take place in the late Universe \cite{Riess:2019cxk,Anderson:2023aga,Wong:2019kwg}, and those from the early Universe where $\Lambda$CDM cosmology is assumed \cite{Aghanim:2018eyx,ACT:2023kun,Schoneberg:2022ggi}. While the appearance of cosmic tensions may be an artifact of some types of measurements \cite{Riess:2020sih,Pesce:2020xfe,deJaeger:2020zpb}, the spectrum across which it appears may call for new physics beyond the $\Lambda$CDM model \cite{Abdalla:2022yfr,Bernal:2016gxb}.

The strength of these challenges has prompted a revaluation of possible alternative formulations of cosmological models with various directions being proposals as the route in which new physics can be incorporated \cite{Sotiriou:2008rp,Nojiri:2010wj,Clifton:2011jh,Bamba:2012cp,Nojiri:2017ncd,CANTATA:2021ktz,Krishnan:2020vaf,Colgain:2022nlb,Malekjani:2023dky,Ren:2022aeo,Dainotti:2021pqg,Addazi:2021xuf,CosmoVerse:2025txj} which have had varying success in explaining the observational data \cite{Schoneberg:2021qvd}. One interesting flavor in which this approach can be embodied is through a modification to the underlying gravitational model which is the most natural since the other components are, largely, determined based on the effect of general relativity in $\Lambda$CDM. These proposals can be preliminarily tested in the local Universe through the combination of supernovae of Type Ia (SNIa), cosmic chronometer (CC), and baryonic acoustic oscillation (BAO) data which have put acute constraints on the cosmological evolution in other settings.

The literature contains a plethora of classes of modified gravity models which are designed to tackle many and varied physical problems \cite{CANTATA:2021ktz,Clifton:2011jh}. General relativity is built on geometric curvature which is the result of a Riemann geometry having being the only mature framework at the time in which the theory was being formulated \cite{e24c2fd0-bd80-346c-b6f7-f15690a270c6}. One approach that has gained traction in recent years has been to consider other geometries, namely torsion in teleparallel gravity \cite{Bahamonde:2021gfp,Aldrovandi:2013wha,Cai:2015emx,Krssak:2018ywd} and nonmetricity in symmetric teleparallel gravity \cite{Nester:1998mp,BeltranJimenez:2019esp}. Metric-affine gravity (MAG) \cite{Hehl:1994ue,Blagojevic:2012bc} extends this even further, and allows all three geometrical entities to be dynamical at the same time. These geometries are embodied in the gravitational connection $\tilde{\Gamma}^{\lambda}\,_{\mu \nu}$ providing the most general geometric framework on which to build gravitational geometrical models where curvature, torsion, and non-metricity are present.

In this study, we focus on the torsion sector within the so-called gauge approach to gravity, where the theory respects Poincaré symmetries and the field strength tensors are the curvature and torsion tensors. This framework naturally leads to a nontrivial hypermomentum tensor, which contains additional intrinsic properties of matter—specifically, intrinsic spin—in addition to the conventional energy-momentum tensor. Although this approach seems natural and well-motivated, its simplest versions, known as quadratic Poincaré theories, are known to propagate ghosts, as the vector and axial components of torsion introduce pathological kinetic terms around generic backgrounds~\cite{BeltranJimenez:2019hrm}. However, it was recently shown in~\cite{Bahamonde:2024sqo} that introducing cubic interactions, which are part of the gauge approach, eliminates these known pathologies and paves the way for a novel framework to study theories with dynamical torsion. In this work, we focus on this theory and study its cosmology within homogeneous and isotropic flat Universe.

In Sec.~\ref{sec:metricaffine}, we briefly introduce the geometrical quantities required to construct theories within Poincaré gauge gravity and present the structure of the cubic theory that is free from Ostrogradsky ghosts in the vector and axial sectors. The cosmological equations of motion are then developed in Sec.~\ref{sec:flrw_eqns} where the flat Friedmann-Lema\^{i}tre-Roberton-Walker (FLRW) metric is considered through a consistency check through an analysis involving the symmetries of the theory. An initial analysis using expansion data is explored in Sec.~\ref{sec:obs_ana} for two branches of the larger framework. Finally, we close with some conclusions and a summary in Sec.~\ref{sec:conclusions}.

\section{Poincaré gauge gravity }\label{sec:metricaffine}

Poincaré Gauge gravity has been suggested as a natural generalization of GR which includes torsion, apart from curvature. In that case though, the covariant derivative of the metric vanishes, which means that the spacetime is metric (its non-metricity is zero). On a more fundamental level, it is directly derived from a gauge theory of gravitation, where the connection of the spacetime plays the role of a gauge field.  

\subsection{Basic ingredients}
Any general linear connection $\tilde{\Gamma}$ that is metric compatible can be decomposed in two parts, 
\begin{equation}\label{general connection}
\tilde{\Gamma}^{\lambda}\,_{\mu \nu}=\Gamma^{\lambda}\,_{\mu \nu}+K^{\lambda}\,_{\mu \nu}
\end{equation}
where $\Gamma ^\lambda {}_{\mu\nu}$ is the Levi-Civita connection and $K^\lambda{}_{\mu\nu}$ is the contortion tensor.
It is necessary to note that, the last  quantitie transform as a tensor, in contrast with the Levi-Civita connection. The expression for the distortion is given by 
\begin{gather}
\Gamma ^\lambda {}_{\mu\nu} = \frac{1}{2}g^{\lambda\kappa}\left(g_{\mu\kappa,\nu} +g_{\kappa\nu,\mu}-g_{\mu\nu,\kappa}\right)\,,\\
K^{\lambda}\,_{\mu \nu}=\frac{1}{2}\left(T^{\lambda}\,_{\mu \nu}-T_{\mu}\,^{\lambda}\,_{\nu}-T_{\nu}\,^{\lambda}\,_{\mu}\right)\,,
\end{gather}
where $T^\lambda{}_{\mu\nu}$ is the antisymmetric part of the connection, i.e. the torsion tensor
\begin{equation}
T^{\lambda}\,_{\mu \nu}=2\tilde{\Gamma}^{\lambda}\,_{[\mu \nu]}\,.
\end{equation}

Then, we can define the general covariant derivative $\tilde{\nabla}_\mu$ with respect to the general metric-compatible connection acting on an arbitrary tensor $T^{\lambda_{1}...\lambda_{i}}\,_{\rho_{1}...\rho_{j}}$ as
\begin{align}
\tilde{\nabla}_\mu T^{\lambda_{1}...\lambda_{i}}\,_{\rho_{1}...\rho_{j}}=&\;\partial_\mu T^{\lambda_{1}...\lambda_{i}}\,_{\rho_{1}...\rho_{j}}+\tilde{\Gamma}^{\lambda_{1}}{}_{\sigma\mu}T^{\sigma...\lambda_{i}}\,_{\rho_{1}...\rho_{j}}+...+\tilde{\Gamma}^{\lambda_{i}}{}_{\sigma\mu_{i}}T^{\lambda_{1}...\sigma}\,_{\rho_{1}...\rho_{j}}\nonumber\\
&-\tilde{\Gamma}^{\sigma}{}_{\rho_{1}\mu}T^{\lambda_{1}...\lambda_{i}}\,_{\sigma...\rho_{j}}-...-\tilde{\Gamma}^{\sigma}{}_{\rho_{j}\mu}T^{\lambda_{1}...\lambda_{i}}\,_{\rho_{1}...\sigma}\,,
\end{align}
and then, they commute as
\begin{equation}
[\tilde{\nabla}_{\mu},\tilde{\nabla}_{\nu}]\,v^{\lambda}=\tilde{R}^{\lambda}\,_{\rho \mu \nu}\,v^{\rho}+T^{\rho}\,_{\mu \nu}\,\tilde{\nabla}_{\rho}v^{\lambda}\,,
\end{equation}
where 
\begin{equation}\label{totalcurvature}
\tilde{R}^{\lambda}\,_{\rho \mu \nu}=\partial_{\mu}\tilde{\Gamma}^{\lambda}\,_{\rho \nu}-\partial_{\nu}\tilde{\Gamma}^{\lambda}\,_{\rho \mu}+\tilde{\Gamma}^{\lambda}\,_{\sigma \mu}\tilde{\Gamma}^{\sigma}\,_{\rho \nu}-\tilde{\Gamma}^{\lambda}\,_{\sigma \nu}\tilde{\Gamma}^{\sigma}\,_{\rho \mu}\,,
\end{equation}
is the curvature tensor. Notice that in the presence of torsion  the curvature tensor does not coincide with the metric (Riemann) tensor, that is defined solely by the Levi-Civita connection. 

The Bianchi identities of the above curvature tensor are generalized as
\begin{gather}
    \tilde{R}^{\lambda}\,_{\mu \nu \rho } = \tilde{\nabla} _{[\mu} T^{\lambda}{}_{\rho\nu]} + T^\sigma {}_{[\mu\rho} T^\lambda {}_{\nu]\sigma}\,,\\
    \tilde{\nabla}_{[\sigma |}\tilde{R}^\lambda {}_{\rho |\mu\nu ]} = T^\kappa {}_{[\sigma \mu |} \tilde{R}^{\lambda} {}_{\rho \kappa | \nu]}\,.
\end{gather}
In the general metric-affine gravity (where nonmetricity is also present), one can define three possible traces of the curvature tensor. However, when nonmetricity is vanishing, there is only one independent trace that is defined as
\begin{equation}
    \tilde{R}_{\mu\nu} = \tilde{R}^\lambda {}_{\mu\lambda\nu}\,, 
\end{equation}
that is the generalised Ricci tensor.  

Regarding the torsion tensor, one can also decompose it into its irreducible parts and thus it can be written as
\begin{equation}\label{irreducibletorsion}
T^{\lambda}\,_{\mu \nu}=\frac{1}{3}\left(\delta^{\lambda}\,_{\nu}T_{\mu}-\delta^{\lambda}\,_{\mu}T_{\nu}\right)+\frac{1}{6}\,\varepsilon^{\lambda}\,_{\rho\mu\nu}S^{\rho}+t^{\lambda}\,_{\mu \nu}\,,
\end{equation}
where
\begin{gather}
T_{\mu}=T^{\nu}\,_{\mu\nu}\,,\\
S_{\mu}=\varepsilon_{\mu\lambda\rho\nu}T^{\lambda\rho\nu}\,,\\
t_{\lambda\mu\nu}=T_{\lambda\mu\nu}-\frac{2}{3}g_{\lambda[\nu}T_{\mu]}-\frac{1}{6}\,\varepsilon_{\lambda\rho\mu\nu}S^{\rho}\,,
\end{gather}
are the torsion vector, axial vector and the torsion tensor (that is a traceless pseudotensor), respectively.

The significance of the above decompositions is well known in the literature. In particular, the axial and vector modes of the torsion tensor play an important role in homogeneous and isotropic cosmological setups, while in such setting the tensor mode becomes trivial \cite{TSAMPARLIS197927}. In addition, the interaction of the axial part of the torsion tensor with Dirac spinors gives interesting phenomenology \cite{Cembranos:2018ipn,Bahamonde:2021gfp}.

\subsection{Gravitational action in Cubic Poincare}\label{sec:action}
Let us now consider the gravitational action of our theory that will be composed by a quadratic contribution $\mathcal{L}^{(2)}$ and a cubic one $\mathcal{L}^{(3)}$. Explicitly, our action reads~\cite{Bahamonde:2024sqo}
\begin{align}
S=&\,\frac{1}{2\kappa^2}\int
\Bigl[
R+\mathcal{L}^{(2)}+\mathcal{L}^{(3)}\Bigr]\sqrt{-g}+\int
\mathcal{L}_{\rm m}\sqrt{-g}\,d^4x\,,\label{Quadratic_Lag}
\end{align}
where $\kappa^2=8\pi G$ and $\mathcal{L}_{\rm m}=\mathcal{L}_{\rm m}(g,\tilde{\Gamma})$ is the matter Lagrangian that in general depends on both the metric and the connection (or equivalently, on the tetrad and the spin connection). 

The quadratic action 
\begin{equation}
    \mathcal{L}^{(2)}=\mathcal{L}^{(2)}_{\rm curv}+\mathcal{L}^{(2)}_{\rm TT}\,, \label{quadLag}
\end{equation} 
contains the following terms\footnote{Note that with respect to~\cite{Bahamonde:2024sqo}, we have a $+R$ in the Lagrangian and our metric signature is $(-+++)$. Then, the mass terms must have a negative sign in front to avoid tachyonic instabilities. Further, the kinetic terms related to the axial and vector sectors must also be negative.}:
\begin{eqnarray}
\mathcal{L}^{(2)}_{\rm curv} &=&   -\frac{1}{2}\left(2c_{1}+c_{2}\right)\tilde{R}_{\lambda\rho\mu\nu}\tilde{R}^{\mu\nu\lambda\rho}+c_{1}\tilde{R}_{\lambda\rho\mu\nu}\tilde{R}^{\lambda\rho\mu\nu}+c_{2}\tilde{R}_{\lambda\rho\mu\nu}\tilde{R}^{\lambda\mu\rho\nu}+d_{1}\tilde{R}_{\mu\nu}\bigl(\tilde{R}^{\mu\nu}-\tilde{R}^{\nu\mu}\bigr)\,,\\
\mathcal{L}^{(2)}_{\rm TT} &=&-\frac{1}{2}m^2_TT_\mu T^\mu-\frac{1}{2}m^2_S S_\mu S^\mu -\frac{1}{2}m^2_tt_{\lambda\mu\nu}t^{\lambda\mu\nu}\,,
\end{eqnarray}
and the cubic Lagrangian $\mathcal{L}^{(3)}$ can be decomposed as:
\begin{equation}
    \mathcal{L}_{\rm}^{(3)}=\mathcal{L}^{(3)}_{\tilde{R}TT}+\mathcal{L}^{(3)}_{\tilde{R}SS}+\mathcal{L}^{(3)}_{\tilde{R}tt}+\mathcal{L}^{(3)}_{\tilde{R}TS}+\mathcal{L}^{(3)}_{\tilde{R}Tt}+\mathcal{L}^{(3)}_{\tilde{R}St}\,,\label{cubicLagIrr}
\end{equation}
where the cubic contribution contains 26 terms
\begin{eqnarray}
    \mathcal{L}^{(3)}_{\tilde{R}TT}&=&h_{1}\tilde{R}_{\mu\nu}T^{\mu}T^{\nu}+h_{2}\tilde{R}T_{\mu}T^{\mu}\,,\\
    \mathcal{L}^{(3)}_{\tilde{R}SS}&=&h_{3}\tilde{R}_{\mu\nu}S^{\mu}S^{\nu}+h_{4}\tilde{R}S_{\mu}S^{\mu}\,,\\
    \mathcal{L}^{(3)}_{\tilde{R}tt}&=&h_5 \tilde{R}_{\lambda\rho\mu\nu}t_{\sigma}{}^{\lambda\rho}t^{\sigma\mu\nu}+h_6\tilde{R}_{\lambda\rho\mu\nu}t_{\sigma}{}^{\lambda\mu}t^{\sigma\rho\nu}+h_7\tilde{R}_{\lambda\rho\mu\nu}t^{\lambda\rho}{}_{\sigma}t^{\sigma\mu\nu}+h_{8}{} \tilde{R}_{\lambda\rho\mu\nu}t^{\lambda\mu}{}_{\sigma}t^{\sigma\rho\nu}+h_{9}{}\tilde{R}_{\lambda\rho\mu\nu}t^{\lambda\mu}{}_{\sigma}t^{\rho\nu\sigma}\nonumber\\
&&+\,h_{10}{}\tilde{R}_{\lambda\rho}t_{\mu\nu}{}^{\lambda}t^{\rho\mu\nu}+h_{11}{}\tilde{R}_{\lambda\rho}t_{\mu\nu}{}^{\lambda}t^{\mu\nu\rho}+h_{12}{}\tilde{R}t_{\lambda\rho\mu}t^{\lambda\rho\mu}\,,\\
    \mathcal{L}^{(3)}_{\tilde{R}TS}&=&h_{13}{}\varepsilon^{\lambda\rho\mu\nu}\tilde{R}_{\lambda\rho\mu\nu}T_{\sigma}S^{\sigma}+h_{14}{} \varepsilon_{\nu}{}^{\lambda\rho\sigma}\tilde{R}_{\lambda\rho\mu\sigma}T^{\mu}S^{\nu}+h_{15}{}\varepsilon^{\lambda\rho\mu\nu}\tilde{R}_{\lambda\rho}T_{\mu}S_{\nu}\,,\\
    \mathcal{L}^{(3)}_{\tilde{R}Tt}&=&h_{16}{}\tilde{R}_{\lambda\rho\mu\nu}T^{\nu}t^{\lambda\rho\mu}+h_{17}{} \tilde{R}_{\lambda\rho\mu\nu}T^{\rho}t^{\lambda\mu\nu}+h_{18}{}\tilde{R}_{\lambda\rho}T_{\mu}t^{\mu\lambda\rho}+h_{19}{} \tilde{R}_{\lambda\rho}T_{\mu}t^{\lambda\rho\mu}\,,\\
    \mathcal{L}^{(3)}_{\tilde{R}St}&=&h_{20}{}\varepsilon_{\alpha\rho\mu\nu}\tilde{R}_{\tau}{}^{\rho\mu\nu}S^{\gamma}t^{\alpha\tau}{}_{\gamma}+h_{21}{}\varepsilon_{\alpha\rho\mu\nu}\tilde{R}_{\tau}{}^{\rho\mu\nu}S^{\gamma}t_{\gamma}{}^{\alpha\tau}+h_{22}{}\varepsilon_{\alpha\rho}{}^{\mu\nu}\tilde{R}^{\rho}{}_{\mu\tau\nu}S^{\gamma}t_{\gamma}{}^{\alpha\tau}+h_{23}{}\varepsilon_{\alpha\rho}{}^{\mu\nu}\tilde{R}_{\gamma\mu\tau\nu}S^{\alpha}t^{\gamma\rho\tau}\nonumber\\
&&+\,h_{24}{}\varepsilon_{\alpha\rho}{}^{\mu\nu}\tilde{R}_{\gamma\mu\tau\nu} S^{\alpha}t^{\rho\tau\gamma}+h_{25}{}\varepsilon_{\alpha\rho\tau\mu}\tilde{R}^{\mu}{}_{\gamma}S^{\alpha}t^{\rho\tau\gamma}+h_{26}{}\varepsilon_{\lambda\rho\mu\nu} \tilde{R}^{\lambda\rho}S_{\sigma}t^{\sigma\mu\nu}\,.
\end{eqnarray}
As shown in~\cite{Bahamonde:2024sqo}, the above theory does not have Ostrogradsky ghosts in the axial and vector sectors for arbitrary backgrounds if one sets:
\begin{eqnarray}\label{condEC}
    h_{2}=-\,\frac{h_{1}}{2}\,,\quad c_{2}=2c_{1}\,, \quad h_{3}= -\,\frac{1}{6}\left(c_{1}+6h_{13}\right)\,,\quad h_{4}= \frac{h_{13}}{2}\,,\quad h_{14}= -\,2h_{13}\,,\quad h_{15}=4h_{13}\,.
\end{eqnarray}
Further, the kinetic terms related to the axial and vector sectors must have the correct signs (negative sign in our convention) leading to
\begin{equation}
    -\,\frac{d_1}{6} \leq c_{1} \leq -\,\frac{d_{1}}{3}\,, \quad d_1 \leq 0\,.\label{kincond}
\end{equation}
Further, the Lagrangian contains quartic terms like $T_\mu T^\mu T^\nu T_\nu, S_\mu S^\mu S^\nu S_\nu, T_\mu S^\mu T^\nu S_\nu$ and $T_\mu T^\mu S^\nu S_\nu$. Then, to avoid that those terms become dominant we require the following inequalities to hold
\begin{equation}
    h_{13} \leq \frac{h_{1}}{16} \leq \bigl(19+6\sqrt{10}\bigr)h_{13}\,, \quad h_{13} < 0\,.\label{quartic}
\end{equation}
In principle, one can include quartic terms in the Lagrangian and avoid the above condition, but for simplicity in our model, we just consider the theory as it was presented in~\cite{Bahamonde:2024sqo}. 

For the matter sector, we can define the variations with respect to the metric as the standard energy-momentum tensor:
\begin{eqnarray}
     T_{\mu\nu} =- \frac{2}{\sqrt{-g}}\frac{\delta (\sqrt{-g}\mathcal{L}_{\rm m})}{\delta g^{\mu\nu}}\,,
\end{eqnarray}
whereas the variations with respect to the connection lead to a new source denoted as hypermomentum defined as
\begin{eqnarray}
  {\Delta_\lambda}^{\mu\nu}=  {-\frac{2}{\sqrt{-g}}} \frac{ \delta(\sqrt{-g}\mathcal{L}_{\rm m})}{\delta \tilde{\Gamma}^\lambda{}_{\mu\nu}}\,\label{hyper}\,,
\end{eqnarray}
that can be decomposed as its spin, dilation, and shear components, but since our theory only contains torsion, it only carries intrinsic spin:
\begin{eqnarray}\label{hypermomentum}
 \Delta_{\mu\nu\lambda}
 ={}^{(\rm s)}\Delta_{[\mu\nu]\lambda}\,.
\end{eqnarray}
Since the gauge approach deals with the pair spin connection and tetrad, we can write down the following equations
\begin{eqnarray}
g_{\mu \nu}&=&e^{a}\,_{\mu}\,e^{b}\,_{\nu}\,g_{a b}\,,\label{vierbein_def}\\
\omega^{a}\,_{b\mu}&=&e^{a}\,_{\lambda}\,e_{b}\,^{\rho}\,\tilde{\Gamma}^{\lambda}\,_{\rho \mu}+e^{a}\,_{\lambda}\,\partial_{\mu}\,e_{b}\,^{\lambda}\,.\label{anholonomic_connection}
\end{eqnarray}
Then, it is also useful to define the canonical energy-momentum tensor that is related to the variations with respect to the tetrad:
\begin{eqnarray}\label{canonicalEM}
\theta_{\mu}\,^{\nu}&=&\frac{e^{a}\,_{\mu}}{\sqrt{- g}}\frac{\delta\left(\sqrt{- g}\mathcal{L}_{\rm m}\right)}{\delta e^{a}\,_{\nu}}\,.
\end{eqnarray}
Note that due to~\eqref{anholonomic_connection}, the hypermomentum tensor defined as~\eqref{hyper} (with respect to the affine connection) coincides with its definition if one varies the matter Lagrangian with respect to the spin connection:
\begin{eqnarray}
   \Delta^{\lambda\mu\nu}&=&-2\frac{e^{a\lambda}e_{b}\,^{\mu}}{\sqrt{- g}}\frac{\delta\left(\sqrt{-g}\mathcal{L}_{\rm m}\right)}{\delta\omega^{a}\,_{b\nu}}= {-\frac{2}{\sqrt{-g}}} \frac{ \delta(\sqrt{-g}\mathcal{L}_{\rm m})}{\delta \tilde{\Gamma}^\lambda{}_{\mu\nu}}\,.
\end{eqnarray}
Having these definitions in mind, we can find the following relationship between these quantities
\begin{eqnarray}
  \Theta{}^{\mu}{}_\lambda=T{}^{\mu}{}_\lambda+\frac{1}{2\sqrt{- g}}\Big[\tilde{\nabla}_\nu (\sqrt{- g}\Delta_\lambda{}^{\mu\nu})-2 \sqrt{- g}\,T_\nu\Delta_\lambda{}^{\mu\nu}\Big]\,,\label{conservation}
\end{eqnarray}
that relates the canonical and metrical energy-momentum tensors with the hypermomentum.
By imposing that the matter sector respects diffeomorphism invariance, we arrive at the following generalised conservation equation~\cite{Hehl:1994ue,Puetzfeld:2014qba,Iosifidis:2020gth}
\begin{eqnarray}
  \sqrt{-g}(2\nabla_{\mu}T^\mu{}_\alpha-\Delta^{\lambda\mu\nu}\tilde{R}_{\lambda\mu\nu\alpha})+\tilde{\nabla}_\mu \tilde{\nabla}_\nu(\sqrt{-g}\Delta_\alpha{}^{\mu\nu})+2T_{\mu\alpha}{}^\lambda \tilde{\nabla}_\nu(\sqrt{-g}\Delta_\lambda{}^{\mu\nu})=0\,.  \label{conserv}
\end{eqnarray}
Note again that $\tilde{\nabla}_\mu$ denotes the general torsionful covariant derivative. The above is the generalization of the energy-momentum conservation law in Metric-Affine geometries. For matter with no microstructure (i.e. $\Delta_{\lambda}{}^{\mu\nu}\equiv 0$) it reduces to the usual conservation law of metric theories.

By taking variations of the action~\eqref{Quadratic_Lag} with respect to the tetrads and spin connection, we arrive at the following set of field equations:
\begin{eqnarray}
 G^{\mu\nu}-\frac{1}{2}( \mathcal{L}^{(2)}+ \mathcal{L}^{(3)}) \,g^{\mu \nu } -\frac{1}{2} V^{\mu \nu } - K^{\mu }{}_{\alpha \kappa } X^{\alpha \nu \kappa } - \nabla_{\alpha}X^{\mu \nu \alpha }   &=&\kappa^2 \theta^{\mu\nu}\,,\label{fieldeq1}\\
\frac{1}{2}\bigl(X^{\lambda \nu \mu }- X^{\lambda \mu \nu }  + X^{\mu \lambda \nu } -  X^{\mu \nu \lambda }\bigr)+2\bigl(K^{\mu }{}_{\alpha \kappa } Y^{\lambda \alpha \nu \kappa } - K^{\lambda }{}_{\alpha \kappa } Y^{\mu \alpha \nu \kappa } - \nabla_{\alpha }Y^{\mu \lambda \nu \alpha }\bigr)&=&\kappa^2 \Delta^{\lambda\mu\nu}\,,\label{fieldeq2}
\end{eqnarray}
where the tensors $X,Y$ and $V$ are given in the appendix~\ref{appendix2}.

\section{FLRW Cosmology in Cubic Poincare gravity}\label{sec:flrw_eqns}
In this section, we study the cosmology of the Poincaré theory presented in the previous section. We first present how the torsion dof can be modelled in FLRW and then, the cosmological equations are presented and studied.
\subsection{Cosmological symmetries}
In our theory, the connection is independent of the metric. Then, we can consider that both quantities satisfy the same symmetries. For doing that, let us take vector fields \(X_{\xi}\), where \(\xi \in \mathfrak{g}\) is an element of the Lie algebra of the symmetry group \(G\). We can define a symmetry under a group action if the Lie derivative of the metric and the connection vanish, i.e.,
\begin{equation}\label{eq:metsymcondi}
(\mathcal{L}_{X_{\xi}}g)_{\mu\nu} = X_{\xi}^{\rho}\partial_{\rho}g_{\mu\nu} + \partial_{\mu}X_{\xi}^{\rho}g_{\rho\nu} + \partial_{\nu}X_{\xi}^{\rho}g_{\mu\rho}=0
\end{equation}
and
\begin{equation}\label{eq:affsymcondi}
\begin{split}
(\mathcal{L}_{X_{\xi}}\Gamma)^{\mu}{}_{\nu\rho} &= X_{\xi}^{\sigma}\partial_{\sigma}\tilde{\Gamma}^{\mu}{}_{\nu\rho} - \partial_{\sigma}X_{\xi}^{\mu}\tilde{\Gamma}^{\sigma}{}_{\nu\rho} + \partial_{\nu}X_{\xi}^{\sigma}\tilde{\Gamma}^{\mu}{}_{\sigma\rho} + \partial_{\rho}X_{\xi}^{\sigma}\tilde{\Gamma}^{\mu}{}_{\nu\sigma} + \partial_{\nu}\partial_{\rho}X_{\xi}^{\mu}\\
&= \tilde{\nabla}_{\rho}\tilde{\nabla}_{\nu}X_{\xi}^{\mu} - X_{\xi}^{\sigma}\tilde{R}^{\mu}{}_{\nu\rho\sigma} - \tilde{\nabla}_{\rho}(X_{\xi}^{\sigma}T^{\mu}{}_{\nu\sigma})=0\,.
\end{split}
\end{equation}
Note that even though the connection is not a tensor, the last condition is a tensorial equation. By assuming cosmological symmetries, the symmetry-generating vector fields are (translations and rotations)
\begin{subequations}\label{eq:genvectrans}
\begin{align}
R_1&=\sin\varphi \partial_\vartheta+\frac{\cos\varphi}{\tan\vartheta}\partial_\varphi\,,\quad R_2=-\cos\varphi \partial_\vartheta+\frac{\sin\varphi}{\tan\vartheta}\partial_\varphi\,,\quad R_3=-\partial_\varphi\,, \\
X_1 &= \chi\sin\vartheta\cos\varphi\partial_r + \frac{\chi}{r}\cos\vartheta\cos\varphi\partial_{\vartheta} - \frac{\chi\sin\varphi}{r\sin\vartheta}\partial_{\varphi}\,,\\
X_2 &= \chi\sin\vartheta\sin\varphi\partial_r + \frac{\chi}{r}\cos\vartheta\sin\varphi\partial_{\vartheta} + \frac{\chi\cos\varphi}{r\sin\vartheta}\partial_{\varphi}\,,\\
X_3 &= \chi\cos\vartheta\partial_r - \frac{\chi}{r}\sin\vartheta\partial_{\vartheta}\,,
\end{align}
\end{subequations}
where \(\chi = \sqrt{1 - kr^2}\) and \(k \in \{-1,0,1\}\) is the sign of the spatial curvature. For simplicity, hereafter, we will only concentrate on the flat ($k=0$) case. By solving the first condition~\eqref{eq:metsymcondi}, the metric becomes the well-known Friedmann-Lemaître-Roberton-Walker (FLRW) metric in Cartesian coordinates which reads
\begin{eqnarray}\label{FLRWmetric}
\dd s^2=-N(t)^2 \dd t^2+a(t)^2(dx^2+dy^2+dz^2)\,,
\end{eqnarray}
where $N(t)$ is the lapse function and $a(t)$ the scale factor of the universe. By assuming that the connection satisfies cosmological symmetries (Eq.~\eqref{eq:affsymcondi} with the vector fields~\eqref{eq:genvectrans}) one can write down the torsion  tensor as \cite{TSAMPARLIS197927}
\begin{align}
   T^\lambda{}_{\mu \nu}&=2T_1\,n_{[\mu}P_{\nu]}{}^{\lambda}+2T_2\,\varepsilon^{\,\lambda}{}_{\mu\nu\rho}n^{\rho} \label{TorsionCos}\,,
\end{align}
where we have split the metric as
\begin{equation}
    g_{\mu\nu}=-n_\mu n_\nu+P_{\mu\nu}\,,
\end{equation}
with $n_\mu$ being a normalised velocity of a cosmological fluid in comoving coordinates, such that $n_\mu=(-N(t),0,0,0)$ satisfying $n_\mu n^\mu=-1$. This means that the torsion tensor contains 2 dof given by $T_i(t)$. The totally antisymmetric tensor $t_{\alpha\mu\nu}$ in the torsion decomposition is vanishing in FLRW cosmology.

Let us now consider the matter sector. Due to the homogeneity and isotropy of a FLRW spacetime,  the energy-momentum tensor is described by the standard perfect fluid form
\begin{align}
    T_{\mu\nu}&=\left(\rho(t)+p(t)\right)n_\mu n_\nu+p(t)g_{\mu\nu}=\rho(t)n_\mu n_\nu+p(t) p_{\mu\nu}\,,\label{energymomentum}
\end{align}
 where $\rho(t)$ and $p(t)$ are the energy density and pressure, respectively. 
 
 Consequently, the canonical energy-momentum tensor~\eqref{canonicalEM} would also acquire the following form
 \begin{align}
    \Theta_{\mu\nu}&=\left(\rho_{\rm C}(t)+p_{\rm C}(t)\right)n_\mu n_\nu+p_{\rm C}(t)g_{\mu\nu}=\rho_{\rm C}(t)n_\mu n_\nu+p_{\rm C}(t) {}_{\mu\nu}\,,\label{Cenergymomentum}
\end{align}
where we have added a subindex ``C" to distinguish between the canonical quantities and the standard ones.

Considering matter described by an unconstrained
hyperfluid respecting  the cosmological principle (isotropy and homogeneity), we find that the hypermomentum is~\cite{Iosifidis:2020gth} 
   \begin{equation}
    	\Delta_{\lambda\mu\nu}=\frac{1}{3}\Delta_1(t) p_{\lambda\mu}n_{\nu}+\Delta_2(t) p_{\lambda\nu}n_{\mu}+\Delta_3(t) n_{\lambda}p_{\mu\nu}+\frac{1}{4}\Delta_4(t) n_{\lambda}n_{\mu} n_{\nu}+\Delta_5(t)\varepsilon_{\lambda\mu\nu\rho}n^{\rho} \label{hyper}\,,
\end{equation}
which contains 5 different sources dof related to the intrinsic spin, dilations, and shears. This can be seen by doing the following decomposition
\begin{eqnarray}
 \Delta_1&=&-\,\frac{3}{4}{}^{(\rm d)}\Delta_4+{}^{(\rm sh)}\Delta_1\,,\quad  \Delta_2=-{}^{(\rm s)}\Delta_3+{}^{(\rm sh)}\Delta_2\,,\quad \Delta_3={}^{(\rm s)}\Delta_3+{}^{(\rm sh)}\Delta_2\,,\\
 \Delta_4&=&{}^{(\rm d)}\Delta_4+4\,{}^{(\rm sh)}\Delta_1\,,\quad \Delta_5={}^{(\rm s)}\Delta_5\,,
\end{eqnarray}
that according to~\eqref{hypermomentum}, we find that since nonmetricity is vanishing, the only non-zero contribution would be related to the intrinsic spin:
\begin{align}
    \Delta_{[\lambda\mu]\nu}&=2\,{}^{(\rm s)}\Delta_3n_{[\lambda}P_{\mu]\nu}+{}^{(\rm s)}\Delta_5\varepsilon_{\lambda\mu\nu\rho}n^\rho\,, \quad {}^{(\rm d)}\Delta_4={}^{(\rm sh)}\Delta_1={}^{(\rm sh)}\Delta_2=0\,.
    \label{hyyper1}
\end{align}
Then, in our Poincare gauge theory, the only non-zero contributions for the hypermomentum are the ones related to the intrinsic spin, i.e., ${}^{(\rm s)}\Delta_3$ and ${}^{(\rm s)}\Delta_5$ and they will act as a source in the connection equations.  Further, since the nonmetricity sources are zero, the Eq.~\eqref{conservation} leads to the following conservation of the hypermomentum tensor
\begin{eqnarray}
    \tilde{\nabla}_\nu (\sqrt{- g}\Delta_\lambda{}^{\mu\nu})=0\,.
\end{eqnarray}
It should be stressed that the above is valid because 
 only the spin part of hypermomentum is switched on. In general, even in the absence of non-metricity the above conservation changes if more parts of hypermomentum are included \cite{Iosifidis:2023but,Dyer:2024kvo}.  Having said that, in our case from Eq.~\eqref{conservation} we notice that the canonical energy-momentum tensor coincides with the standard energy-momentum tensor leading to $\rho=\rho_{\rm C}$ and $p=p_{\rm C}$.

\subsection{Flat FLRW cosmological equations}
By assuming the cosmological symmetries in flat FLRW as explained in the previous section, we notice that the quadratic Lagrangian~\eqref{quadLag} has contributions from $c_1,c_2$ and the masses $m_S,m_T$. Further, the cubic Lagrangian~\eqref{cubicLagIrr} only has contributions from $h_1, h_2, h_3, h_4, h_{13}$ and $h_{14}$. Explicitly, these Lagrangians in flat FLRW become
\begin{eqnarray}
    \mathcal{L}_2&=&-\frac{12 c_1 }{N^2}\left(\dot{T}_2-N T_2 \left(H-2 T_1\right)\right){}^2+\frac{6 c_2 }{N^2}\left(2 N T_2 \left(5 H-4 T_1\right) \dot{T}_2+N^2 T_2{}^2 \left(-16 H T_1+13 H^2+4 T_1{}^2\right)+\dot{T}_2{}^2\right)\nonumber\\
    &&+\frac{6\dot{H}}{N}+12 H^2+72 m_S^2 T_2{}^2+\frac{9}{2} m_T^2 T_1{}^2\,,\\
     \mathcal{L}_3&=&\frac{27 h_1 T_1{}^2 }{N}\left(-\dot{H}+H N \left(T_1-H\right)+\dot{T}_1\right)-\frac{54 h_2 T_1{}^2 }{N}\left(\dot{H}+N \left(-3 H T_1+2 H^2+T_1{}^2-T_2{}^2\right)-\dot{T}_1\right)\nonumber\\
     &&-\frac{432 h_3 T_2{}^2 }{N}\left(\dot{H}+H N \left(H-T_1\right)-\dot{T}_1\right)+\frac{864 h_4 T_2{}^2 }{N}\left(-\dot{H}+N \left(3 H T_1-2 H^2-T_1{}^2+T_2{}^2\right)+\dot{T}_1\right)\nonumber\\
     &&-\frac{432 h_{13} T_2 T_1 }{N}\left(N T_2 \left(3 H-2 T_1\right)+\dot{T}_2\right)-\frac{216 h_{14} T_2 T_1 }{N}\left(H N T_2+\dot{T}_2\right)\,,
\end{eqnarray}
where dots denote derivative with respect to time $t$ and $H=\dot{a}/(aN)$ is the Hubble parameter.
Hereafter, we will assume the stability conditions~\eqref{condEC} to avoid Ostrogradsky ghosts in the axial and vector sectors. One may compute the cosmological equations either directly from the above Lagrangian or from the tensorial field equations. Upon setting $N(t)=1$ and using~\eqref{fieldeq1}, the modified flat FLRW equations take the form:
\begin{eqnarray}
  \kappa^2 \rho&=&3H^2+\frac{9}{2}m_T^2 \left(\frac{ T_1^2}{2}-H T_1\right)+36 m_S^2 T_2^2+H^2 \left(\frac{135}{2} h_{1} T_1^2+216 h_{13} T_2^2\right)+H \Big[27 (h_{1}-16 h_{13}) T_1 T_2^2-54 h_{1} T_1^3\Big]\nonumber\\
  &&-27 h_{1} H^3 T_1-\frac{27}{2} (h_{1}-16 h_{13}) T_1^2 T_2^2+\frac{27}{2} h_{1} T_1^4+216 h_{13} T_2^4\,,\label{FRW1}\\
   -\kappa^2p   &=&3H^2+2\dot{H}+m_T^2 \left[-3 H T_1-\frac{3 \dot{T}_1}{2}+\frac{3 T_1^2}{4}\right]+12 m_S^2 T_2^2+\dot{H} \left(18 h_{1} T_1^2+144 h_{13} T_2^2\right)-18 h_{1} H^3 T_1\nonumber\\
   &&+H^2 \Big[-9 h_{1} \dot{T}_1+\frac{63}{2} h_{1} T_1^2+360 h_{13} T_2^2\Big]+\dot{T}_1 \left[(9 h_{1}-144 h_{13}) T_2^2-27 h_{1} T_1^2\right]+(18 h_{1}-288 h_{13}) T_1 T_2 \dot{T}_2\nonumber\\
    &&+H \Big[-18 h_{1} T_1 \dot{H}+(18 h_{1}-288 h_{13}) T_1 T_2^2+36 h_{1} T_1 \dot{T}_1-9 h_{1} T_1^3+288 h_{13} T_2 \dot{T}_2\Big]+\left(\frac{9 h_{1}}{2}-72 h_{13}\right) T_1^2 T_2^2\nonumber \\
   &&-\frac{1}{2} 9 h_{1} T_1^4-72 h_{13} T_2^4\,,\label{FRW2}
\end{eqnarray}
whereas the FLRW equations obtained from the connection field equations~\eqref{fieldeq2} yield
\begin{eqnarray}
  2   \kappa ^2 \,{}^{(\rm s)}\Delta_3&=&3 T_1 \Big[6h_1 (H-T_1) (H-2 T_1)-6 (h_{1}-16 h_{13}) T_2^2+m_T^2\Big]\,,\label{FRW3}\\
   \kappa ^2 \,{}^{(\rm s)}\Delta_5&=&3 T_2 \Big[48 h_{13} \left(-H^2+T_1^2+2 T_2^2\right)-3h_1 T_1^2+8 m_S^2\Big]\,.\label{FRW4}
\end{eqnarray}
By using the conservation equation~\eqref{conserv} we find:
\begin{eqnarray}
    \dot{\rho}+3H(\rho+p)=3  \,{}^{(\rm s)}\Delta_5\Big(H T_2+\dot{T}_2\Big)-3  \,{}^{(\rm s)}\Delta_3 \left(\dot{H}-H T_1+H^2-\dot{T}_1\right)\,.\label{conserv1}
\end{eqnarray}
The right hand side of this equation provides new corrections to the standard conservation equation that depends on the hypermomentum. This equation can be alternatively found by combining the generalised FLRW equations~\eqref{FRW1} and~\eqref{FRW2} and using the connection FLRW equations~\eqref{FRW3} and~\eqref{FRW4}. Further, by using the connection field equations, we can rewrite the conservation equation as
\begin{eqnarray}
    \dot{\rho}+3H(\rho+p)&=&-\frac{9}{2 \kappa ^2} \Big[-H \Big\{T_1{}^2 \Big(18 h_1 (\dot{H}-\dot{T}_1)-12 (h_1-16 h_{13}) T_2{}^2+m_T^2\Big)+16 T_2{}^2 \Big(12 h_{13} T_2{}^2+m_S^2\Big)+12 h_1 T_1{}^4\Big\}\nonumber\\
    &&+H^2 \Big\{T_1 \Big(6 h_1 \dot{H}-6 h_1 \dot{T}_1-6 \left(h_1-16 h_{13}\right) T_2{}^2+m_T^2\Big)+96 h_{13} T_2 \dot{T}_2+30 h_1 T_1{}^3\Big\}+6 h_1 H^4 T_1\nonumber\\
    &&+T_1 \Big(12 h_1 T_1{}^2-6 \left(h_1-16 h_{13}\right) T_2{}^2+m_T^2\Big) \Big(\dot{H}-\dot{T}_1\Big)-24 H^3 \Big(h_1 T_1{}^2-4 h_{13} T_2{}^2\Big)\nonumber\\
    &&-2 T_2 \dot{T}_2 \Big(-3 \left(h_1-16 h_{13}\right) T_1{}^2+96 h_{13} T_2{}^2+8 m_S^2\Big)\Big]\,.\label{conserv2}
\end{eqnarray}
As it is obvious,  the right hand side of this equation vanishes when hypermomentum is zero since the connection equations must hold.

\section{Data Sets and Model Comparisons}

As we mentioned before, in this work, we consider three sets of observational data; the latest Pantheon+ compilation, Cosmic Chronometers and DESI 1yr BAO measurements. In this section we discuss in short the data sets we used and we present the statistical measures with which we compared the models. 

\subsection{Data Sets}

Pantheon+ consists of SNIa observations \cite{Brout:2021mpj,Riess:2021jrx,Scolnic:2021amr} which contains 1701 light curves of 1550 unique spectroscopically confirmed SNIa that span up to redshift $z=2.3$. Specifically, the data set contains measurements of apparent magnitudes $m$, each accompanied by corresponding statistical uncertainties, compiled across a range of redshift values. It further includes a covariance matrix that encodes systematic uncertainties and inter-measurement correlations. Recent analyses utilizing the Pantheon+ compilation have identified potential issues, including an estimated $5\%$ overstatement of the uncertainties encoded in the covariance matrix \cite{Keeley:2022iba}, as well as the presence of a previously unmodeled systematic - volumentric redshift scatter - which may indicate a genuine evolution in the absolute magnitude of Type Ia supernovae at a redshift of $z = 0.005$ \cite{Perivolaropoulos:2023iqj}.

Using the observed apparent magnitude, $m$, and the absolute magnitude, $M_B$, SNIa can be used to determine the Hubble rate, $H(z)$, using
\begin{equation}
    \mu (z_i,\Theta) = m - M_B = 5 \log _{10}[D_L(z_i,\Theta)]+25\,,
\end{equation}
where $z_i$ is the redshift of the SNIa measurement, $D_L(z_i,\Theta)$ is the luminosity distance given by
\begin{equation}
    D_L(z_i,\Theta) = c(1+z_i) \int _0 ^{z_i} \frac{dz'}{H(z',\Theta)}\,,
\end{equation}
and $\Theta$ are collectively the set of cosmological parameters. In order to constrain the cosmological parameters, we use $\chi ^2$ which reads
\begin{equation}
    \chi _{\rm SN} ^2 = (\mu (z_i,\Theta)-\mu_{\rm obs} (z_i))^T C_{\rm SN}^{-1} (\mu (z_i,\Theta)-\mu_{\rm obs} (z_i))\,,
\end{equation}
with $C_{\rm SN}$ is the covariance matrix of the data set.

In addition, we consider the latest data set of CC $H(z)$ measurements, together with the full covariance matrix which contains calibration and systematic errors \cite{Moresco:2020fbm}. This data set is in a sense model independent since it does not assume a particular cosmological model \textit{a priori}, but depends on the differential age technique between galaxies, spanning the redshift range up to $z \sim 2.$ Though an emerging model-independent probe of the Universe's expansion history, the CC approach remains subject to ongoing scrutiny regarding its robustness and reliability within the cosmological community \cite{Abdalla:2022yfr}. Major observational programs - such as SDSS, DESI and DES - tend to prioritize alternative probes like Type Ia Supernovae (SNIa), Baryon Acoustic Oscillations (BAO), and galaxy clustering due to their comparatively higher statistical weight and maturity. The CC technique, which infers the Hubble parameter $H(z)$ from differential age measurements of galaxies, inherently depends on the integrated cosmic history and thus suffers from limited statistical precision and pronounced sensitivity to systematic effects. These include uncertainties in the calibration of Stellar Population Synthesis (SPS) models and challenges in constructing clean CC samples free from contaminating populations \cite{Moresco:2023zys}. While CC data are included in our analysis, their interpretative power is diminished by systematics related to stellar metallicity, biases stemming from assumptions about star formation histories, and inaccuracies in spectral modeling \cite{Moresco:2022phi}. Such limitations may introduce non-negligible biases, complicating the extraction of cosmological parameters and reducing the method's standalone reliability (see Refs. \cite{Mukherjee:2020vkx,Gomez-Valent:2018hwc,Haridasu:2018gqm} for assessments of CC-related systematics). Consequently, our results rest on the provisional assumption of the CC data's validity, while emphasizing the necessity for further methodological refinement and empirical validation. 

In our MCMC analysis, we evaluated the discrepancy of the observational Hubble data values $H_{\rm obs} (z_i)$, from the theoretical Hubble parameter values $H(z_i,\Theta)$ with model parameters $\Theta$, with an observational error of $\sigma _H (z_i)$, using $\chi _{\rm CC}^2$ which is calculated through
\begin{equation}
    \chi _{\rm CC} ^2 = (H (z_i,\Theta)-H_{\rm obs} (z_i))^T C_{\rm CC}^{-1} (H (z_i,\Theta)-H_{\rm obs} (z_i))\,,
\end{equation}

Furthermore, we used the DESI first year data sets (DR1) of BAO measurements \cite{DESI:2024mwx,DESI:2024lzq,DESI:2024uvr,DESI:2024aqx,DESI:2016fyo,DESI:2024kob}, derived from observations of tens of millions of extragalactic sources, including galaxies, quasars and Lyman-$\alpha$ forest tracers. These data sets offer high-precision constraints on the transverse comoving distance, the Hubble expansion rate and their angle-averaged combination- each expressed relative to the comoving sound horizon at the drag epoch - across seven redshift bins spanning the range $0.1 < z < 4.2$. DR1 represents a major milestone in observational cosmology. Based on 13 months of data collection, it includes high-confidence spectroscopic redshifts for approximately 18.7 million sources, making it the most extensive dataset of its kind to date. In addition to BAO, DR1 also includes full-shape analyses of the two-point correlation function, allowing for complementary constraints on the growth of structure. Together, these results yield tight bounds on $\Lambda$CDM parameters, including $\Omega _{\rm m} \simeq 0.296$ and $H_0 \simeq 68.6$ km/s/Mpc, and place competitive limits on the sum of neutrino masses. Initial findings also show mild tension with the standard cosmological constant models, hinting at possible time variation in dark energy. In order to avoid overlapping data sets, we consider only DESI data and decide not to include other BAO data sets for this reason.

\subsection{Model Comparison} \label{sec:mod_com}
To assess the performance of each selected model across different datasets, we employ standard statistical measures. The primary metric used is the minimum chi-squared value, $\chi^2_{\mathrm{min}}$, which is derived from the maximum likelihood via the relation $\chi^2_{\mathrm{min}} = -2 \ln \mathcal{L}_{\mathrm{max}}$, as outlined in the previous sections. A lower $\chi^2_{\mathrm{min}}$ indicates a better fit between the model and the observational data.

Beyond evaluating goodness of fit through $\chi^2_{\mathrm{min}}$, we further compare each model to the standard cosmological model, $\Lambda$CDM, using two model selection criteria: the Akaike Information Criterion (AIC) and the Bayesian Information Criterion (BIC). These criteria not only account for how well a model fits the data, but also penalize model complexity based on the number of free parameters, $n$. The AIC is defined as
\begin{equation}
\mathrm{AIC} = \chi^2_{\mathrm{min}} + 2n \,,
\end{equation}
where lower values indicate a more favourable balance between goodness of fit and model simplicity. Although a model may achieve a lower $\chi^2_{\mathrm{min}}$, the AIC penalizes excessive parameterisation to discourage over-fitting. Similarly, the BIC is given by
\begin{equation}
\mathrm{BIC} = \chi^2_{\mathrm{min}} + n \ln m \,,
\end{equation}
where $m$ denotes the total number of data points in the combined observational dataset. Like the AIC, the BIC seeks to balance model fit and complexity but imposes a stricter penalty on models with more parameters, particularly as the sample size increases.

To facilitate comparison across models, we compute the differences in AIC and BIC relative to the $\Lambda$CDM model:
\begin{align} \label{eq:aic_delta}
\Delta \mathrm{AIC}&= \mathrm{AIC}_{\text{model}} - \mathrm{AIC}_{\Lambda \text{CDM}} = \Delta \chi^2_{\textrm{min}} + 2\Delta n \,,\\
\label{eq:bic_delta}
\Delta \mathrm{BIC} &= \mathrm{BIC}_{\text{model}} - \mathrm{BIC}_{\Lambda \text{CDM}} = \Delta \chi^2_{\textrm{min}} + \Delta n \ln m \,.
\end{align}
These quantities quantify how each model deviates from the reference model. Smaller values of $\Delta \mathrm{AIC}$ and $\Delta \mathrm{BIC}$ indicate models that are more strongly supported by the data, while also offering a suitable balance between fit quality and model complexity.

The MCMC posteriors, along with the results for the standard cosmological model, are presented in Appendix.~\ref{sec:lcdm}.

\section{Cosmological analysis}\label{sec:obs_ana}

In this section, we explore two distinct scenarios arising from the cosmological implications of our model. These scenarios correspond to different assumptions about the behavior and role of hypermomentum in the evolution of the Universe. First, we consider the case where hypermomentum contributions vanish, leading to dynamics governed solely by the standard perfect fluid sector. Then, we analyze the scenario where the perfect fluid and hypermomentum sectors evolve independently, allowing for richer phenomenology due to nontrivial torsional effects. As for the data set combinations, we use CC as our benchmark given that it contains $H(z)$ data directly, while considering SN and DESI data sets separately and then combined together.

\subsection{Vanishing Hypermomentum \texorpdfstring{${}^{(\rm s)}\Delta_3={}^{(\rm s)}\Delta_5=0$}{}}\label{sec:branch_1}
In this section, we will assume that hypermomentum corrections are very small and hence, one can neglect their effects. This means that we assume that the microstructure of matter does not affect the dynamics of cosmology and matter is fully described by the standard perfect fluid sector. This means that the conservation equation~\eqref{conserv1} becomes the standard one with its right hand side not affecting the dynamics of the fluid. 

In this case, the connection FLRW equations take the form of algebraic equations for \(T_1\) and \(T_2\). These equations can be solved in four different ways for arbitrary values of the constants \(h_i\). Substituting these solutions into the FLRW equations reveals that the system's dynamics split into two distinct branches. However, setting \(h_1 = 0\) reduces these branches to a single one, simplifying the dynamics of the model to depend solely on the parameters \(m_S\) and \(h_{13}\). Note that one cannot choose $h_{13}=0$ without setting $h_1=0$ due to the condition~\eqref{quartic}. To note, this is also the limit in which this branch becomes $\Lambda$CDM.

\begin{figure}[htbp]
    \centering
    \includegraphics[scale = 0.5]{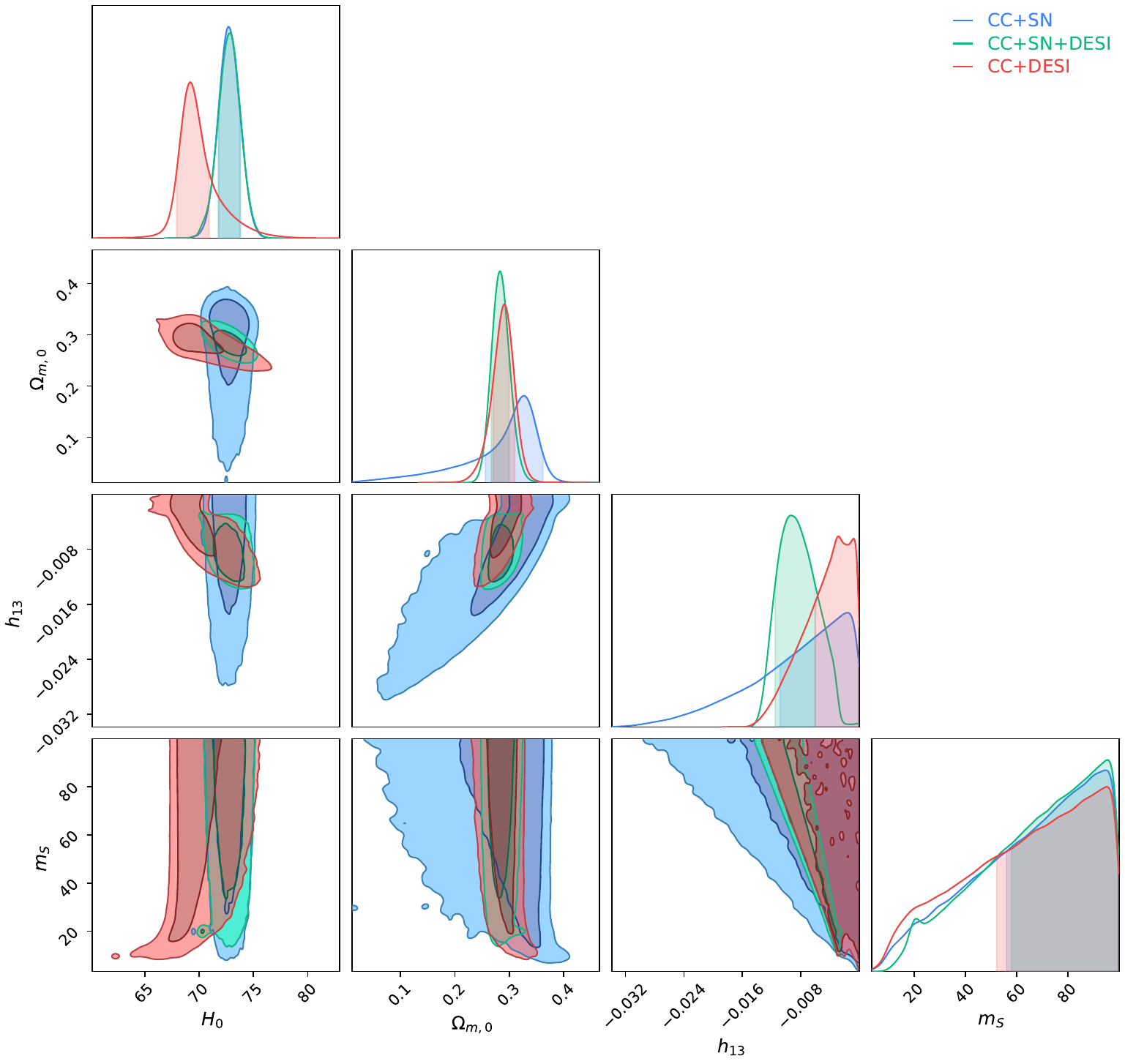}
    \caption{Confidence contours and posterior distributions for the Vanishing Hypermomentum model parameters $h_{13}$ and $m_S$ and also including $H_0$ and $\Omega_{m,0}$. The blue contours represent the CC+SN dataset, the red contours illustrate CC+DESI, while the green contours correspond to the combination CC+SN+DESI.}
\label{fig:branch_1}

\end{figure}

Therefore, the free parameters in our MCMC analyses are the Hubble constant $H_0$, the present-day matter density parameter $\Omega_{m,0}$, and the model-specific parameters $h_{13}$ and $m_S$. The posterior distributions of these parameters are shown in Fig.~\ref{fig:branch_1}, which show fairly Gaussian behaviour except for the $m_S$ parameter which suffers a steep decline at very high values. These posteriors are derived using three different dataset combinations: CC+SN, CC+SN+DESI, and CC+DESI. The corresponding numerical results are presented in Table~\ref{tab:model1_outputs}, where the mass $m_S$ is found to have very high values showing super-Planckian in terms of their field values.

The most prominent effect observed in the posterior distributions is the influence of the supernova data, which shifts the inferred value of $H_0$ to higher values. Another notable feature—especially when all three datasets are combined—is the anti-correlation between $H_0$ and $\Omega_{m,0}$. This behaviour is physically expected, as the matter density decreases with the expansion of the Universe. In addition, the parameter $M$, which encodes the absolute magnitude of Type Ia SN, is found to be consistent with values reported in the literature.

Since $h_1$ has been set to zero in this case, the remaining model parameters are $h_{13}$ and $m_S$. $h_{13}$ controls the strength of the coupling between the curvature and the torsion vectors. $m_{S}$ is the mass associated to the torsion vector. The posterior distributions illustrate how these two parameters correlate with one another and with the cosmological parameters. A slight anti-correlation is observed between $h_{13}$ and $m_S$. More notably, there is a clear positive correlation between the matter density $\Omega_{m,0}$ and $h_{13}$. Additionally, the parameter $m_S$ tends to favour values towards the upper end of its allowed range. In contrast, $h_{13}$ is centred around zero, with its 1$\sigma$ confidence interval extending primarily into the negative domain. However, in both cases, the values remain consistent with those of $\Lambda$CDM at $2\sigma$ level.

Table~\ref{tab:model1_stats} presents the values of $\chi^2_{\mathrm{min}}$, along with the corresponding AIC and BIC values and their differences relative to the $\Lambda$CDM model, as described in Sec.~\ref{sec:mod_com}. Naturally, the lowest absolute values of $\chi^2_{\mathrm{min}}$, AIC, and BIC are associated with the CC+DESI dataset; however, this is primarily due to the smaller size of this dataset compared to, for example, the SN dataset.

The most informative results are found in the final two columns of Table~\ref{tab:model1_stats}, which report the differences in AIC and BIC relative to the standard $\Lambda$CDM model, as shown in Sec.~\ref{sec:lcdm}. These values provide insight into the relative support for each model given the data. In the cases of the CC+SN and CC+DESI combinations, the slightly positive values of $\Delta$AIC and $\Delta$BIC suggest a mild preference for the $\Lambda$CDM model, as these models exhibit either similar or marginally poorer performance after accounting for model complexity.

The most compelling result arises with the combined dataset CC+SN+DESI. In this case, both $\Delta$AIC and $\Delta$BIC are negative, meaning that the proposed model achieves better fit statistics than $\Lambda$CDM even though it has a higher parameter count. Negative values in these criteria indicate that the alternative model is slightly more supported by the data, even after penalizing for complexity. This suggests that when all datasets are considered together, the proposed model may offer a slightly more statistical sound description of the Universe compared to the standard cosmological model. Such a result may point toward the potential for this model to be studied further.

\begin{table}[h]
    \centering
    \label{tab:model1_outputs}
    \begin{tabular}{cccccc}
        \hline
        \rule{0pt}{3ex}
		Model & $H_0\, [{\rm km\, s}^{-1} {\rm Mpc}^{-1}]$ & $\Omega_{m,0}$ & $M$ & $h_{13}$ & $m_S [M_{\rm pl}]$ \\ [4 pt]
		\hline
        \rule{0pt}{4ex}
        CC+SN & $72.68^{+1.05}_{-0.90}$ & $0.329^{+0.032}_{-0.072}$ & $-19.265^{+0.032}_{-0.025}$ & $\left( -7.8^{+6.9}_{-109.3} \right) \times 10^{-4}$ & $96.8^{+3.0}_{-43.4}$  \\ [4 pt]
		CC+SN+DESI & $72.85^{+0.97}_{-1.05}$ & $0.283^{+0.016}_{-0.015}$ & $-19.269^{+0.030}_{-0.029}$ & $\left( -9.4^{+3.4}_{-2.1} \right) \times 10^{-3}$ & $96.2^{+3.6}_{-41.0}$ \\[4 pt] 
		CC+DESI & $69.1^{+1.8}_{-1.1}$ & $0.291^{+0.019}_{-0.021}$ & -- & $\left( -2.9^{+2.8}_{-3.6} \right) \times 10^{-3}$ & $96.8^{+2.9}_{-47.8}$ \\[4 pt] 

		\hline

    \end{tabular}
    \caption{Exact results for the Vanishing Hypermomentum model, including the parameters $H_0$, $\Omega_{m,0}$, $h_{13}$ and $m_S$, for different dataset combinations. }
\end{table}

\begin{table}[h]
\centering
    \label{tab:model1_stats}
\begin{tabular}{l c c  c || c c}
\hline
\rule{0pt}{3ex}
 Data Sets & $\chi^2_{\mathrm{min}}$& AIC & BIC & $\Delta$AIC & $\Delta$BIC \\[4 pt] 
		\hline
        \rule{0pt}{4ex}
CC+SN & 1539.22 & 1549.22 & 1555.39 & 4.00 & 1.00\\[4 pt]
CC+SN+DESI & 1555.22 & 1565.16 & 1571.34 & -7.85 & -5.67\\[4 pt]
CC+DESI & 19.08 & 29.08 & 26.24 & 3.99 & 2.86 \\[4 pt]
\hline
\end{tabular}
\caption{Comparison of $\chi^2_{min}$ min and differences in AIC and BIC between the Vanishing Hypermomentum model and $\Lambda$CDM (i.e $\Delta$AIC and $\Delta$BIC).}
\end{table}

\subsection{Perfect fluid sector and hypermomentum are conserved independently}\label{sec:branch_2}

Let us now study the limiting case when the conservation equations for the fluid and hypermomentum are conserved independently. For non-zero hypermomentum, this means that the torsional functions become (see Eq.~\eqref{conserv1})
\begin{eqnarray}
    T_1(t)=\frac{K_1}{a(t)}+H\,,\quad T_2(t)=\frac{K_2}{a(t)}\,.
\end{eqnarray} 

Then, each energy density component behaves  as $\rho\propto a^{-3(1+w)}$ with $w=\{1/3,0,w\}$ for 
radiation, matter, and dark energy, respectively. In this case, the hypermomentum functions are fully determined by the connection equations:
\begin{eqnarray}
   2  \,{}^{(\rm s)}\Delta_3a^3&=& 3 (a H+K_1) \left(6 h_1 K_1 a H+m_T^2 a^2+12 h_1 K_1^2-6 h_1 K_2^2+96 h_{13} K_2^2\right)\,,\\
   \,{}^{(\rm s)}\Delta_5a^3&=& 3 K_2 \left(-6 K_1 a H (h_1-16 h_{13})+a^2 \left(8 m_S^2-3 h_1 H^2\right)-3 h_1 K_1^2+48 h_{13} \left(K_1^2+2 K_2^2\right)\right)\,,
\end{eqnarray}
and the remaining FLRW equation that dictates the dynamic of the scale factor is
\begin{eqnarray}
  \frac{\rho_{\rm m0}}{a^3}+\frac{\rho_{\rm r0}}{a^4}+\rho_{\rm de0} a^{-3 (w+1)}  &=&\frac{1}{4 a^4}\Big[9 a^2 \left(6 h_1 H^2 \left(K_2^2-K_1^2\right)+K_1^2 m_T^2+16 K_2^2 m_S^2\right)+3 \left(4-3 m_T^2\right) a^4 H^2\nonumber\\
  &&+54 h_1 \left(K_1^4-K_1^2 K_2^2\right)+864 h_{13} K_2^2 \left(K_1^2+K_2^2\right)\Big]\,.
\end{eqnarray}
Setting 
\begin{equation}
    C_{1}=\frac{27}{2}h_{1}(K_2^2-K_1^2) \;\;, \;\; C_{2}=\frac{9}{4}(K_1^2 m_T^2+16 K_2^2 m_S^2)\;\;, \;\; C_{3}=\frac{3}{4}(4-3 m_{T}^{2})
\end{equation}
\begin{equation}
    C_{4}=\frac{1}{2}\Big[ 27 h_{1}(K_1^4-K_1^2 K_2^2)+432 h_{13} K_2^2 (K_1^2+K_2^2)\Big]
\end{equation}
we may express the latter as
\begin{equation}
    \Big(C_{3}+\frac{C_{1}}{a^{2}} \Big)H^{2}= \frac{\rho_{\rm m0}}{a^3}+\frac{(\rho_{\rm r0}-C_{4})}{a^4}+\frac{\rho_{\rm de0} }{a^{3 (w+1)}}-\frac{C_{2}}{a^{2}}
\end{equation}
where each of the $C_{i}'s$ has a clear physical meaning, $C_{1}$ and $C_{3}$ modify the gravitational coupling, $C_{2}$ is an effective 'spatial curvature' term and $C_{4}$ alters the radiation energy density, all coming due to the presence of hypermomentum. 

In this scenario, we allow $C_1$ and $C_2$ to vary freely, while fixing $C_3 = 1$. The latter acts merely as a normalization constant for the gravitational sector and does not restrict additional deviations, which can still be captured through $C_1$. We also set $C_4$ such that the total effective radiation term matches the standard one. The $\Lambda$CDM limit occurs for the setting where $C_1$and $C_2$ tends to zero.

\begin{figure}[htbp]
    \centering
    \includegraphics[scale = 0.5]{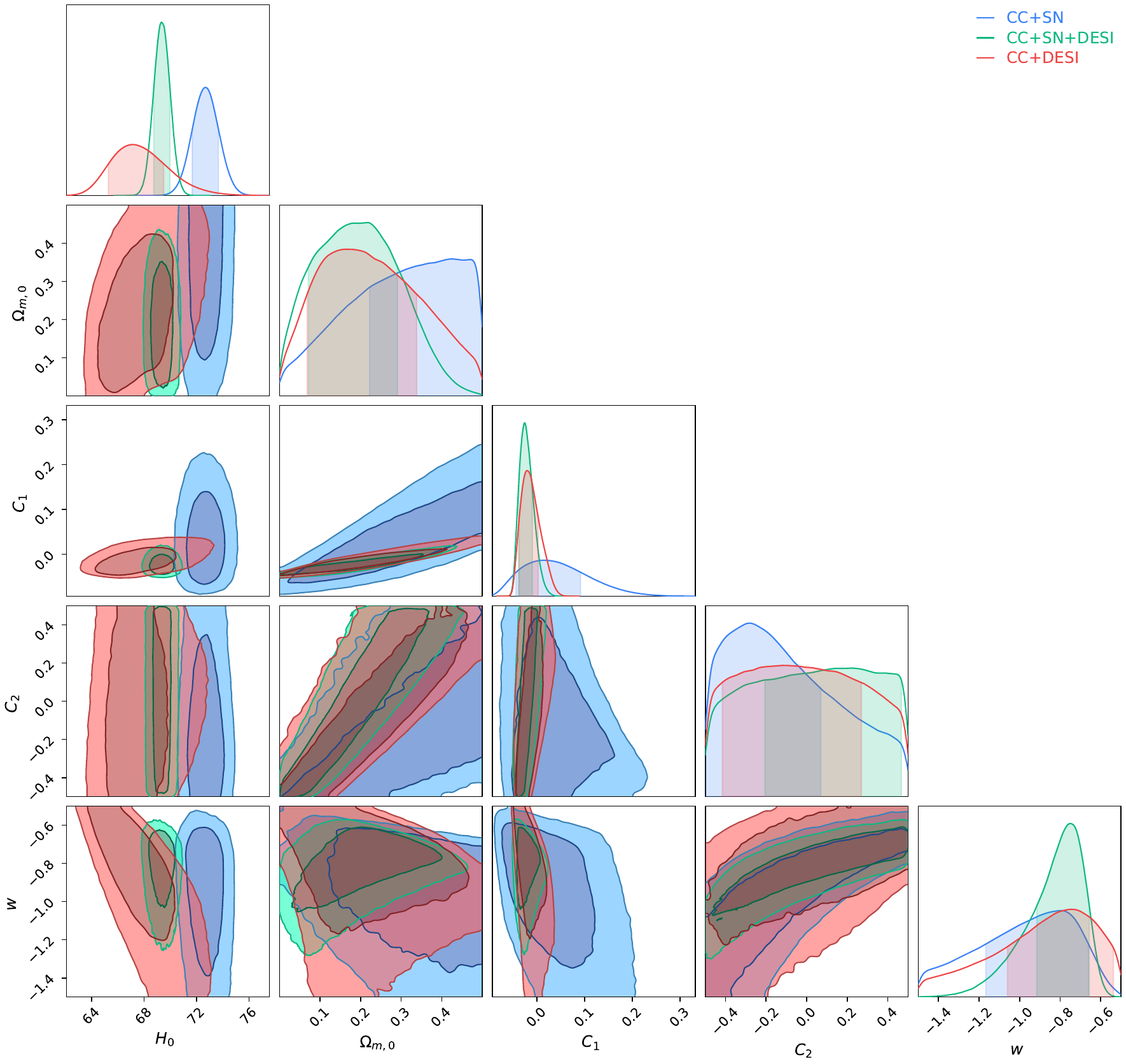}
    \caption{Confidence contours and posterior distributions for the Independent conservation of fluid and hypermomentum model parameters  which include $C_1$, $C_2$ and $w$. $H_0$ and $\Omega_{m,0}$ are also constrained. The blue contours represent the CC+SN dataset, the red contours illustrate CC+DESI, while the green contours correspond to the combination CC+SN+DESI.}
\label{fig:branch_2}
\end{figure}

Therefore, the free parameters in this setup are $H_0$, $\Omega_{m,0}$ (associated with the matter energy density), the model parameters $C_1$ and $C_2$, and the equation of state parameter $w$, which remains independent due to the separate conservation of the fluid and hypermomentum components. In addition, the nuisance parameter $M$ is included in the MCMC analysis. The resulting posterior distributions and parameter correlations are shown in Fig.~\ref{fig:branch_2}. As expected, the inclusion of the SN dataset shifts the preferred $H_0$ toward higher values, while the CC+DESI data favours slightly lower values. The combined dataset yields constraints that lie between these two tendencies.

We observe that $C_1$ is tightly constrained around zero, indicating minimal deviations from the standard gravitational coupling in this term. In contrast, $C_2$ exhibits broader constraints and extends into the negative region within the $1\sigma$ confidence level. The equation-of-state parameter $w$ is well-constrained in the negative region and remains close to the cosmological constant value of $w = -1$. Although not exactly equal to $-1$, this value consistently lies within the $1\sigma$ confidence interval. A clear correlation between $w$ and $C_2$ is also visible in the joint posterior contours.

The exact numerical results are reported in Table~\ref{tab:model2_outputs}, which also includes the inferred values of the nuisance parameter $M$. Across all dataset combinations, $w$ is found to be around $-0.75$, suggesting an equation of state  slightly different from a pure cosmological constant.  The corresponding statistical measures for each dataset combination are presented in Table~\ref{tab:branch_2_stats}.

In this case, we observe that the standard $\Lambda$CDM model is slightly preferred across all combinations, as both the $\Delta$AIC and $\Delta$BIC values are positive—implying that the information criteria for $\Lambda$CDM are lower. However, it is important to maintain perspective: this extended model naturally incurs a penalty due to its larger number of free parameters. Although the statistical preference leans toward $\Lambda$CDM, the moderate values of $\Delta$AIC and $\Delta$BIC suggest that the extended model is not strongly disfavoured and remains a viable alternative.

\begin{table}
    \centering
    \label{tab:model2_outputs}
    \begin{tabular}{ccccccc}
        \hline
        \rule{0pt}{3ex}
		Model & $H_0\, [{\rm km\, s}^{-1} {\rm Mpc}^{-1}]$ & $\Omega_{m,0}$ & $M$ & $C_1$ & $C_2$ & $w$ \\ 
		\hline
        \rule{0pt}{4ex}
		CC+SN & $72.70^{+0.97}_{-1.02}$ & $0.50^{+0.00}_{-0.28}$ & $-19.258^{+0.028}_{-0.030}$ & $0.013^{+0.079}_{-0.057}$ & $-0.28^{+0.35}_{-0.22}$ & $-0.79^{+0.13}_{-0.38}$ \\ [4 pt]
		CC+SN+DESI & $69.32^{+0.65}_{-0.59}$ & $0.216^{+0.075}_{-0.145}$ & $-19.367\pm 0.016$ & $-0.026^{+0.016}_{-0.013}$ & $0.22^{+0.25}_{-0.42}$ &  $-0.752^{+0.095}_{-0.164}$  \\ [4 pt]
		CC+DESI & $67.1^{+2.4}_{-1.8}$ & $0.159^{+0.179}_{-0.091}$ & -- & $-0.021^{+0.024}_{-0.016}$ & $-0.12^{+0.39}_{-0.30}$ & $-0.74^{+0.20}_{-0.32}$ \\ [4 pt]
		\hline
    \end{tabular}
\caption{ Exact results for the second model, including the parameters $H_0$, $\Omega_{m,0}$, $C_1$, $C_2$, and $w$ for different dataset combinations.}
\end{table}

\begin{table}[h]
\centering
    \label{tab:branch_2_stats}
\begin{tabular}{l c c  c || c c}
\hline
\rule{0pt}{3ex}
Data Sets & $\chi^2_{\textrm{min}}$ & AIC & BIC & $\Delta$AIC & $\Delta$BIC\\
\hline
\rule{0pt}{4ex}
CC+SN & 1538.15& 1550.14 & 1557.55 & 4.92 & 9.35 \\ [4 pt]
CC+SN+DESI & 1569.44 & 1581.44 & 1588.87 & 8.15 & 11.86 \\ [4 pt]
CC+DESI &17.77 & 29.77 & 26.35 & 4.69 & 2.99 \\ [4 pt]
\hline
\end{tabular}
\caption{Comparison of $\chi^2_{min}$ min and differences in AIC and BIC between the second model and $\Lambda$CDM (i.e $\Delta$AIC and $\Delta$BIC).}
\end{table}

\section{Conclusions}\label{sec:conclusions}
The analysis pursued in this work builds on the extensive work in the literature establishing the foundations of Poincaré gauge gravity and the subclass of cubic models, and extends this to a first investigation of how specific models within this class compare with observational measurements in the late Universe. The work opens with a brief introduction on Poincaré gauge gravity where the full geometric generality of the underlying theory is explored through a general linear connection $\tilde{\Gamma}^{\lambda}\,_{\mu \nu}$ that is however compatible with the metric. This connection is used to build geometric objects that span the full breadth of possible Riemann-Cartan frameworks. We probe this space of models up to cubic order in Eq.~\eqref{Quadratic_Lag} where stability and Ostrogradsky ghost conditions are incorporated into the expression of the action through a combination of a priori requirements and conditions discussed below this definition. In principle, one may also include cubic curvature invariants (31 possible terms) and fix the theory so as to avoid ghosts. This can be achieved in a manner similar to what has been done in the Riemannian sector in Refs.~\cite{Erices:2019mkd,Asimakis:2022mbe}. Another possible extension of the model is the inclusion of quartic torsional terms (33 possible terms), analogous to those appearing in certain invariants constructed from the torsional Gauss–Bonnet term~\cite{Kofinas:2014owa}. In the present work, we have neglected such terms; however, it would be interesting to investigate their effects in future studies.

The equations of motion for this formulation of cubic Poincaré gauge gravity for a flat FLRW geometry are expressed in Sec.~\ref{sec:flrw_eqns}. Here, the symmetries of the geometry are first used to determine the conserved quantities of the system. This is directly followed by equations of motion for the system variables, as well as the fluid conservation equation. This produces a system of coupled differential equations that need to be resolved through the consideration of subclasses of the overarching theory. Indeed, in Sec.~\ref{sec:obs_ana}, two initial branches are considered for this extensive framework theory. In the first branch, the hypermomentum,  which is the response of the matter action under connection variations, vanishes in order to explore the role of the cubic interaction term $h_{13}$ and the mass term of the pseudovector $m_S$. The mass term $m_S$ is consistent with $\Lambda$CDM but has a preference for a nonzero value in all cases at a level of at least $2\sigma$, while the $h_{13}$ parameter also produces a slight preference for a nonzero value. This shows some viability beyond $\Lambda$CDM and may benefit from further analysis with other datasets probing other aspects of cosmology, including further data on large scale structure as well as CMB measurements from the early Universe.

For the second branch in Sec.~\ref{sec:obs_ana}, the matter fluid and hypermomentum equations are restricted to be conserved independently. This provides another physically interesting branch of the overarching class of theories contained in the cubic Poincaré gauge gravity framework. This is particularly interesting since the background cosmology equations provide a framework in which a flat FLRW geometry produces the same term as a nonflat geometry in terms of the dependence on the scale factor. This gives another route to introduce that term without the need to induce nonflat geometries. In other words, torsional effects induce or mimic a curvature-like term in the flat-FLRW modified equations. It should be noted that at the background level, $C_2$ contributes as $a^{-2}$ and therefore enters $H(z)$ in the same way as a true curvature density parameter $\Omega_k$. However, the two cases are not fully degenerate. In a genuinely curved FLRW spacetime, curvature also affects the relation between comoving distances and observables through $D_M(z)=S_k[\chi(z)]$, with $S_k$ the curvature-dependent function. In contrast, in our flat model with a $C_2/a^2$ term the metric stays spatially flat, so the transverse comoving distance is simply $D_M(z)=\chi(z)$ with $\chi(z)=\int_0^z \mathrm{d}z'/H(z')$~\cite{Weinberg:2008zzc}. Thus, probes relying only on the expansion history (radial BAO, cosmic chronometers, etc.) would be unable to distinguish $C_2$ from $\Omega_k$, while geometric consistency tests that compare transverse and radial distances (e.g.\ CMB angular-diameter distance, transverse BAO, Alcock--Paczynski tests) can break this degeneracy~\cite{Aghanim:2018eyx,Eisenstein:2005su,2dFGRS:2005yhx,Alcock:1979mp,Seo:2003pu}. 

The effect of the DESI data set is more pronounced in the second branch, which has the impact of reducing the Hubble constant value. While the cosmology parameter best-fits are in agreement with $\Lambda$CDM at an appropriate statistical level, the value of the equation of state does appear to be generally lower.

It would be interesting to explore scenarios where hypermomentum and the cosmic fluid are not conserved independently, giving rise to interacting models in which the intrinsic spin modifies the standard conservation equation of the cosmic fluid. Such systems would naturally be more complex and merit further investigation. Moreover, a recently constructed cubic theory that includes nonmetricity without Ostrogradsky ghosts in the vector/axial sectors~\cite{Bahamonde:2024efl} provides a natural extension for our framework. Another important avenue for future work is to study cosmological perturbations following~\cite{Aoki:2023sum} within this general class of models. 
In our framework, both the metric $g_{\mu\nu}$ and the torsion tensor $T^\lambda{}_{\mu\nu}$ are dynamical fields, and therefore, a complete perturbative analysis requires decomposing not only the metric but also the torsion tensor, which carries 24 independent components. 
Performing such a decomposition and the subsequent scalar–vector–tensor analysis is a highly nontrivial task, especially given the presence of cubic interactions in our Poincaré gauge theory. It is natural to expect that not all torsional dof are dynamical in a ghost-free theory. In this sense, new healthy vector dof can propagate around an FLRW background in this theory, which is not possible in the standard quadratic Poincaré theory~\cite{Yo:2001sy,Yo:1999ex,BeltranJimenez:2019hrm}. Furthermore, the parity-preserving quadratic healthy theory possesses only an extra scalar dof, and only one independent torsional function appears at the background level~\cite{Shie:2008ms}. This feature does not allow one to obtain dynamics such as those described in Sec.~\ref{sec:branch_2}, where hypermomentum acts as an effective spatial curvature term in the flat FLRW cosmological equations.
For these reasons, we first focused on the background cosmology to establish the main physical features of the model. 
Building on this foundation, future work will address the full cosmological perturbation analysis, the underlying stability and viability of the theory, and a comprehensive confrontation with cosmological datasets sensitive to these structures.

\noindent
\section*{Acknowledgments}
This paper is based upon work from COST Action CA21136 {\it Addressing observational tensions in cosmology with systematics and fundamental physics} (CosmoVerse) supported by COST (European Cooperation in Science and Technology). S.B. is supported by “Agencia Nacional de Investigación y Desarrollo” (ANID), Grant “Becas Chile postdoctorado al extranjero” No. 74220006 and also by the Institute for Basic Science under the project code IBS-R018-D3. JLS would also like to acknowledge funding from ``The Malta Council for Science and Technology'' as part of the ``FUSION R\&I: Research Excellence Programme'' REP-2023-019 (CosmoLearn) Project. JLS and RB would also like to acknowledge funding from ``Xjenza Malta'' as part of the ``Technology Development Programme'' DTP-2024-014 (CosmicLearning) Project. K.F.D. was supported by the PNRR-III-C9-2022–I9 call, with project number 760016/27.01.2023. D.I.  acknowledges the support of  Istituto Nazionale di Fisica Nucleare (INFN), Sezioni  di Napoli ,  {\it Iniziative Specifiche} QGSKY.

\newpage
\appendix

\section{Field equations}\label{appendix2}
The definition of the tensors appearing in the spin connection equation~\eqref{fieldeq2} is:
\begin{align}
X^{\lambda\mu\nu}&=U^{\lambda[\mu\nu]}\,,\quad Y^{\lambda\rho\mu\nu}=W^{[\lambda\rho][\mu\nu]}\,,
\end{align}
where
\begin{align}
    U_{\lambda \rho \mu } =&\;-\frac{1}{2} m_{T}^2 \big(g_{\lambda \rho } T_{\mu } -  g_{\lambda \mu } T_{\rho }\big)+2 m_{S}^2 \big(T_{\lambda \mu \rho } +  T_{\mu \rho\lambda} + T_{\rho \lambda \mu }\big) - \frac{1}{3} m_{t}^2 \big(2 T_{\lambda \mu \rho } + T_{\mu \lambda \rho } +  T_{\rho\mu \lambda} + g_{\lambda \mu } T_{\rho }-  g_{\lambda \rho } T_{\mu }\big)\nonumber\\
    &+\bar{h}_{1}^{} \big(\tilde{R}^{\alpha \kappa }{}_{\mu \rho } + \tilde{R}_{\mu \rho }{}^{\alpha \kappa }\big) T_{\lambda \alpha \kappa } + 
2 \bar{h}_{2}^{} \tilde{R}_{\rho }{}^{\kappa }{}_{\mu }{}^{\alpha } T_{\lambda \alpha \kappa } + 
\bar{h}_{3}^{} \big(\tilde{R}^{\alpha \kappa }{}_{\mu \rho } T_{\alpha \lambda \kappa } + \tilde{R}_{\lambda \rho }{}^{\alpha \kappa } T_{\mu \alpha \kappa }\big)\nonumber\\
& - 
\bar{h}_{4}^{} \big(\tilde{R}_{\rho }{}^{\alpha }{}_{\mu }{}^{\kappa } T_{\alpha \lambda \kappa } +  \tilde{R}_{\lambda }{}^{\kappa }{}_{\rho }{}^{\alpha } T_{\mu \alpha \kappa }\big)- 
\bar{h}_{5}^{} \big( \tilde{R}_{\rho }{}^{\kappa }{}_{\mu }{}^{\alpha } T_{\alpha \lambda \kappa } +  \tilde{R}_{\rho }{}^{\kappa }{}_{\lambda }{}^{\alpha } T_{\mu \alpha \kappa }\big) + 
\bar{h}_{6}^{} \big(\tilde{R}_{\mu \rho }{}^{\alpha \kappa } T_{\alpha \lambda \kappa } + \tilde{R}^{\alpha \kappa }{}_{\lambda \rho } T_{\mu \alpha \kappa }\big) \nonumber\\
&+ 
\bar{h}_{7}^{} \big(g_{\lambda \mu } \tilde{R}_{\rho }{}^{\alpha \kappa \theta } T_{\alpha \theta \kappa } -  \tilde{R}_{\lambda }{}^{\alpha }{}_{\mu \rho } T_{\alpha }\big) + 
\bar{h}_{8}^{} \big( \tilde{R}_{\rho }{}^{\alpha }{}_{\lambda \mu } T_{\alpha }- g_{\lambda \mu } \tilde{R}_{\rho }{}^{\theta \alpha \kappa } T_{\alpha \theta \kappa }\big) + 
\bar{h}_{9}^{} \big( \tilde{R}_{\lambda \rho \mu }{}^{\alpha } T_{\alpha }- g_{\lambda \mu } \tilde{R}^{\alpha \theta }{}_{\rho }{}^{\kappa } T_{\alpha \theta \kappa }\big)\nonumber\\
& + 
\bar{h}_{10}^{} \big(g_{\lambda \mu } \tilde{R}^{\kappa \theta }{}_{\rho }{}^{\alpha } T_{\alpha \theta \kappa } -  \tilde{R}_{\mu \rho \lambda }{}^{\alpha } T_{\alpha }\big) + 
\bar{h}_{11}^{} \big(\tilde{R}^{\alpha \kappa }{}_{\lambda \rho } + \tilde{R}_{\lambda \rho }{}^{\alpha \kappa }\big) T_{\alpha \mu \kappa } + 
2 \bar{h}_{12}^{} \tilde{R}_{\lambda }{}^{\alpha }{}_{\rho }{}^{\kappa } T_{\alpha \mu \kappa } \nonumber\\
&+ 
\bar{h}_{13}^{} \big(\tilde{R}_{\lambda }{}^{\kappa }{}_{\rho }{}^{\alpha } + \tilde{R}_{\rho }{}^{\alpha }{}_{\lambda }{}^{\kappa }\big) T_{\alpha \mu \kappa } + 
2 \bar{h}_{14}^{} \tilde{R}_{\rho }{}^{\kappa }{}_{\lambda }{}^{\alpha } T_{\alpha \mu \kappa }+ 
\bar{h}_{15}^{} \big(\tilde{R}_{\alpha \lambda } + \tilde{R}_{\lambda \alpha }\big) T^{\alpha }{}_{\mu \rho } + 
\bar{h}_{16}^{} \big( \tilde{R}_{\alpha \lambda } T_{\rho \mu }{}^{\alpha }- \tilde{R}_{\rho \alpha } T^{\alpha }{}_{\lambda \mu }\big)  \nonumber\\
&+ 
\bar{h}_{17}^{} \big( \tilde{R}_{\lambda \alpha } T_{\rho \mu }{}^{\alpha }- \tilde{R}_{\alpha \rho } T^{\alpha }{}_{\lambda \mu }\big) + 
\bar{h}_{18}^{} \big(\tilde{R}_{\alpha \rho } + \tilde{R}_{\rho \alpha }\big) T_{\lambda \mu }{}^{\alpha } + 
\bar{h}_{19}^{} \big(\tilde{R}_{\alpha \rho } + \tilde{R}_{\rho \alpha }\big) T_{\mu \lambda }{}^{\alpha } 
- \bar{h}_{20}^{} g_{\lambda \mu } \big(\tilde{R}_{\alpha \rho } + \tilde{R}_{\rho \alpha }\big) T^{\alpha }\nonumber\\
& + 
\bar{h}_{21}^{} \big(g_{\lambda \mu } \tilde{R}_{\alpha \kappa } T^{\alpha }{}_{\rho }{}^{\kappa } -  \tilde{R}_{\lambda \rho } T_{\mu }\big) + 
\bar{h}_{22}^{} \big(g_{\lambda \mu } \tilde{R}_{\alpha \kappa } T^{\kappa }{}_{\rho }{}^{\alpha } -  \tilde{R}_{\rho \lambda } T_{\mu }\big) - 
\bar{h}_{23}^{} \big(g_{\lambda \mu } \tilde{R}_{\alpha \kappa } T_{\rho }{}^{\alpha \kappa } +  \tilde{R}_{\rho \mu } T_{\lambda }\big)\nonumber\\
& + 
2 \bar{h}_{24}^{} \tilde{R} T_{\lambda \mu \rho } + 
\bar{h}_{25}^{} \tilde{R} \big(T_{\mu \lambda \rho } -  T_{\rho \lambda \mu }\big) + 
\bar{h}_{26}^{} \tilde{R} \big(g_{\lambda \rho } T_{\mu } -  g_{\lambda \mu } T_{\rho }\big) \,,\\
     W_{\lambda \rho \mu \nu } =&\;2 c_{1}^{} \tilde{R}_{\rho \lambda \nu \mu } - 2 c_{2}^{} \tilde{R}_{\nu \rho \lambda \mu }- (2 c_{1}^{} + c_{2}^{}) \tilde{R}_{\nu \mu \rho \lambda } + d_{1}^{} \big[g_{\nu \rho } \big(\tilde{R}_{\lambda \mu } -  \tilde{R}_{\mu \lambda }\big) + g_{\lambda \nu } \big(\tilde{R}_{\mu \rho } -  \tilde{R}_{\rho \mu }\big)\big]\nonumber\\
    &+\bar{h}_{1}^{} T_{\alpha \lambda \rho } T^{\alpha }{}_{\mu \nu } 
- \bar{h}_{2}^{} T_{\alpha \lambda \mu } T^{\alpha }{}_{\nu \rho }  
- \bar{h}_{3}^{} T_{\alpha \mu \nu } T_{\lambda \rho }{}^{\alpha } + 
\bar{h}_{4}^{} T_{\alpha \nu \rho } T_{\lambda \mu }{}^{\alpha } + 
\bar{h}_{5}^{} T_{\alpha \nu \rho } T_{\mu \lambda }{}^{\alpha } 
- \bar{h}_{6}^{} T_{\alpha \lambda \rho } T_{\mu \nu }{}^{\alpha }\nonumber\\
    & + 
\bar{h}_{7}^{} T_{\rho \mu \nu } T_{\lambda } 
- \bar{h}_{8}^{} T_{\mu \nu \rho } T_{\lambda } 
- \bar{h}_{9}^{} T_{\lambda \nu \rho } T_{\mu } + 
\bar{h}_{10}^{} T_{\nu \lambda \rho } T_{\mu } + 
\bar{h}_{11}^{} T_{\lambda \rho }{}^{\alpha } T_{\mu \nu \alpha } + 
\bar{h}_{12}^{} T_{\lambda \mu }{}^{\alpha } T_{\rho \nu \alpha } + 
\bar{h}_{13}^{} T_{\mu \lambda }{}^{\alpha } T_{\rho \nu \alpha } \nonumber\\
    &+ 
\bar{h}_{14}^{} T_{\mu \lambda }{}^{\alpha } T_{\nu \rho \alpha } + 
\bar{h}_{15}^{} g_{\nu \rho } T_{\lambda }{}^{\alpha }{}_{\kappa } T_{\mu \alpha }{}^{\kappa } + 
\bar{h}_{16}^{} g_{\nu \rho } T_{\alpha \lambda }{}^{\kappa } T_{\mu }{}^{\alpha }{}_{\kappa } + 
\bar{h}_{17}^{} g_{\nu \rho } T_{\alpha \mu }{}^{\kappa } T_{\lambda }{}^{\alpha }{}_{\kappa } + 
\bar{h}_{18}^{} g_{\nu \rho } T_{\alpha \lambda }{}^{\kappa } T^{\alpha }{}_{\mu \kappa }\nonumber\\
    & + 
\bar{h}_{19}^{} g_{\nu \rho } T_{\alpha \lambda }{}^{\kappa } T_{\kappa \mu }{}^{\alpha } + 
\bar{h}_{20}^{} g_{\nu \rho } T_{\lambda } T_{\mu } + 
\bar{h}_{21}^{} g_{\nu \rho } T_{\lambda \mu }{}^{\alpha } T_{\alpha } + 
\bar{h}_{22}^{} g_{\nu \rho } T_{\mu \lambda }{}^{\alpha } T_{\alpha } + 
\bar{h}_{23}^{} g_{\nu \rho } T_{\alpha \lambda \mu } T^{\alpha } \nonumber\\
    &+ 
\bar{h}_{24}^{} g_{\lambda \mu } g_{\nu \rho } T_{\alpha }{}^{\kappa \theta } T^{\alpha }{}_{\kappa \theta } + 
\bar{h}_{25}^{} g_{\lambda \mu } g_{\nu \rho } T_{\alpha }{}^{\kappa \theta } T_{\kappa }{}^{\alpha }{}_{\theta } + 
\bar{h}_{26}^{} g_{\lambda \mu } g_{\nu \rho } T_{\alpha } T^{\alpha }\,,
\end{align}
and the definition of the tensor $V^{\mu\nu}$ appearing in the tetrad equation~\eqref{fieldeq1} is
\begin{align}
V_{\mu \nu}=&\; d_{1}^{} \big(2 \tilde{R}_{\nu }{}^{\kappa }{}_{\mu }{}^{\alpha } \tilde{R}_{\alpha \kappa } -2 \tilde{R}_{\nu }{}^{\alpha }{}_{\mu }{}^{\kappa } \tilde{R}_{\alpha \kappa } - 2 \tilde{R}_{\alpha \mu } \tilde{R}^{\alpha }{}_{\nu } + 2 \tilde{R}^{\alpha }{}_{\mu } \tilde{R}_{\nu \alpha }\big)+4 c_{1}^{} \tilde{R}_{\theta \kappa \mu \alpha } \big(\tilde{R}^{\kappa \theta }{}_{\nu }{}^{\alpha } -  \tilde{R}_{\nu }{}^{\alpha \kappa \theta }\big)  \nonumber\\
&+ 2 c_{2}^{} \tilde{R}_{\theta \kappa \mu \alpha } \big(\tilde{R}^{\alpha \theta }{}_{\nu }{}^{\kappa } -  \tilde{R}_{\nu }{}^{\alpha \kappa \theta } + \tilde{R}_{\nu }{}^{\theta \kappa \alpha }\big)  + 4 m_{S}^2 \big(T_{\alpha \nu }{}^{\kappa } T^{\alpha }{}_{\mu \kappa } -  T^{\alpha }{}_{\mu \kappa } T^{\kappa }{}_{\nu \alpha } + T^{\kappa }{}_{\mu \alpha } T_{\nu }{}^{\alpha }{}_{\kappa }\big) \nonumber\\
&+ m_{T}^2 \big(T_{\nu \mu }{}^{\alpha } T_{\alpha } -  T_{\mu } T_{\nu }\big) -  \frac{2}{3} m_{t}^2 \big(2 T_{\alpha \nu }{}^{\kappa } T^{\alpha }{}_{\mu \kappa } + T^{\alpha }{}_{\mu \kappa } T^{\kappa }{}_{\nu \alpha } -  T^{\kappa }{}_{\mu \alpha } T_{\nu }{}^{\alpha }{}_{\kappa } + T_{\nu \mu }{}^{\alpha } T_{\alpha } -  T_{\mu } T_{\nu }\big) \nonumber\\
&-2 \bar{h}_{1}^{} \big(\tilde{R}_{\theta }{}^{\tau }{}_{\nu \kappa } T_{\alpha \mu }{}^{\kappa } -  \tilde{R}_{\nu \kappa }{}^{\tau }{}_{\theta } T_{\alpha \mu }{}^{\kappa } + \tilde{R}_{\theta }{}^{\tau }{}_{\mu \kappa } T_{\alpha \nu }{}^{\kappa }\big) T^{\alpha \theta }{}_{\tau } 
-2 \bar{h}_{2}^{} \big(\tilde{R}_{\kappa }{}^{\tau }{}_{\nu \theta } T_{\alpha \mu }{}^{\kappa } -  \tilde{R}_{\nu }{}^{\tau }{}_{\kappa \theta } T_{\alpha \mu }{}^{\kappa } + \tilde{R}_{\kappa }{}^{\tau }{}_{\mu \theta } T_{\alpha \nu }{}^{\kappa }\big) T^{\alpha \theta }{}_{\tau } \nonumber\\
&+ 
\bar{h}_{3}^{} \big(\tilde{R}_{\nu }{}^{\alpha \tau }{}_{\theta } T_{\alpha \mu }{}^{\kappa } T_{\kappa }{}^{\theta }{}_{\tau } + 2 \tilde{R}^{\theta \tau }{}_{\nu \kappa } T_{\alpha \mu }{}^{\kappa } T_{\tau }{}^{\alpha }{}_{\theta } + 2 \tilde{R}^{\theta \tau }{}_{\mu \kappa } T_{\alpha \nu }{}^{\kappa } T_{\tau }{}^{\alpha }{}_{\theta } -  \tilde{R}^{\tau }{}_{\theta \alpha \kappa } T_{\tau \mu }{}^{\theta } T_{\nu }{}^{\alpha \kappa }\big) \nonumber\\
&+ 
\bar{h}_{4}^{} \big(\tilde{R}_{\kappa }{}^{\tau }{}_{\nu }{}^{\theta } T_{\alpha \mu }{}^{\kappa } T_{\tau }{}^{\alpha }{}_{\theta }- \tilde{R}^{\alpha \tau }{}_{\nu \theta } T_{\alpha \mu }{}^{\kappa } T_{\kappa }{}^{\theta }{}_{\tau } -  \tilde{R}^{\alpha \tau }{}_{\mu \theta } T_{\alpha \nu }{}^{\kappa } T_{\kappa }{}^{\theta }{}_{\tau } \nonumber\\
&-  \tilde{R}_{\nu }{}^{\tau }{}_{\kappa }{}^{\theta } T_{\alpha \mu }{}^{\kappa } T_{\tau }{}^{\alpha }{}_{\theta } + \tilde{R}_{\kappa }{}^{\tau }{}_{\mu }{}^{\theta } T_{\alpha \nu }{}^{\kappa } T_{\tau }{}^{\alpha }{}_{\theta } + \tilde{R}_{\kappa }{}^{\tau }{}_{\alpha \theta } T_{\tau \mu }{}^{\theta } T_{\nu }{}^{\alpha \kappa }\big) \nonumber\\
&+ 
\bar{h}_{5}^{} \big(\tilde{R}_{\kappa }{}^{\theta }{}_{\nu }{}^{\tau } T_{\alpha \mu }{}^{\kappa } T_{\tau }{}^{\alpha }{}_{\theta }- \tilde{R}_{\nu }{}^{\tau \alpha }{}_{\theta } T_{\alpha \mu }{}^{\kappa } T_{\kappa }{}^{\theta }{}_{\tau } -  \tilde{R}_{\nu }{}^{\theta }{}_{\kappa }{}^{\tau } T_{\alpha \mu }{}^{\kappa } T_{\tau }{}^{\alpha }{}_{\theta }  \nonumber\\
&+ \tilde{R}_{\kappa }{}^{\theta }{}_{\mu }{}^{\tau } T_{\alpha \nu }{}^{\kappa } T_{\tau }{}^{\alpha }{}_{\theta } + \tilde{R}_{\kappa }{}^{\tau }{}_{\mu \theta } T_{\alpha }{}^{\theta }{}_{\tau } T_{\nu }{}^{\alpha \kappa } + \tilde{R}_{\kappa \theta \alpha }{}^{\tau } T_{\tau \mu }{}^{\theta } T_{\nu }{}^{\alpha \kappa }\big) \nonumber\\
&+ 
\bar{h}_{6}^{} \big(2 \tilde{R}_{\nu \kappa }{}^{\theta \tau } T_{\alpha \mu }{}^{\kappa } T_{\tau }{}^{\alpha }{}_{\theta }- \tilde{R}_{\theta }{}^{\tau }{}_{\nu }{}^{\alpha } T_{\alpha \mu }{}^{\kappa } T_{\kappa }{}^{\theta }{}_{\tau } -  \tilde{R}_{\theta }{}^{\tau }{}_{\mu }{}^{\alpha } T_{\alpha \nu }{}^{\kappa } T_{\kappa }{}^{\theta }{}_{\tau } -  \tilde{R}_{\theta }{}^{\tau }{}_{\mu \kappa } T_{\alpha }{}^{\theta }{}_{\tau } T_{\nu }{}^{\alpha \kappa } -  \tilde{R}_{\alpha \kappa }{}^{\tau }{}_{\theta } T_{\tau \mu }{}^{\theta } T_{\nu }{}^{\alpha \kappa }\big) \nonumber\\
&+ 
\bar{h}_{7}^{} \big[\tilde{R}_{\alpha }{}^{\kappa }{}_{\tau }{}^{\theta } T_{\kappa }{}^{\tau }{}_{\theta } T_{\nu \mu }{}^{\alpha }+ \tilde{R}_{\nu }{}^{\alpha }{}_{\theta \kappa } T_{\alpha }{}^{\kappa \theta } T_{\mu } - 2 (\tilde{R}_{\alpha }{}^{\theta }{}_{\nu \kappa } T_{\theta \mu }{}^{\kappa } + \tilde{R}_{\alpha }{}^{\theta }{}_{\mu \kappa } T_{\theta \nu }{}^{\kappa }) T^{\alpha }\big]\nonumber\\
&+ 
\bar{h}_{8}^{} \big[\tilde{R}_{\alpha \tau }{}^{\kappa \theta } T_{\kappa }{}^{\tau }{}_{\theta } T_{\nu \mu }{}^{\alpha } -  \tilde{R}_{\nu \kappa }{}^{\alpha }{}_{\theta } T_{\alpha }{}^{\kappa \theta } T_{\mu }-  \big(\tilde{R}_{\alpha \kappa \nu }{}^{\theta } T_{\theta \mu }{}^{\kappa } -  \tilde{R}_{\nu \alpha }{}^{\theta }{}_{\kappa } T_{\theta \mu }{}^{\kappa } + \tilde{R}_{\alpha \kappa \mu }{}^{\theta } T_{\theta \nu }{}^{\kappa } + \tilde{R}_{\alpha \kappa \mu \theta } T_{\nu }{}^{\kappa \theta }\big) T^{\alpha }\big] \nonumber\\
&+ 
\bar{h}_{9}^{} \big(\tilde{R}^{\kappa }{}_{\tau \alpha }{}^{\theta } T_{\kappa }{}^{\tau }{}_{\theta } T_{\nu \mu }{}^{\alpha } + \tilde{R}_{\kappa }{}^{\theta }{}_{\nu \alpha } T_{\theta \mu }{}^{\kappa } T^{\alpha } + \tilde{R}_{\nu }{}^{\theta }{}_{\alpha \kappa } T_{\theta \mu }{}^{\kappa } T^{\alpha } + \tilde{R}_{\kappa }{}^{\theta }{}_{\mu \alpha } T_{\theta \nu }{}^{\kappa } T^{\alpha } -  \tilde{R}^{\alpha }{}_{\kappa \nu \theta } T_{\alpha }{}^{\kappa \theta } T_{\mu } -  \tilde{R}^{\alpha }{}_{\kappa \mu \theta } T_{\alpha }{}^{\kappa \theta } T_{\nu }\big) \nonumber\\
&+ 
\bar{h}_{10}^{} \big(\tilde{R}_{\tau }{}^{\theta }{}_{\alpha }{}^{\kappa } T_{\kappa }{}^{\tau }{}_{\theta } T_{\nu \mu }{}^{\alpha } - 2 \tilde{R}_{\nu \kappa \alpha }{}^{\theta } T_{\theta \mu }{}^{\kappa } T^{\alpha } -  \tilde{R}_{\theta \kappa \mu \alpha } T_{\nu }{}^{\kappa \theta } T^{\alpha } + \tilde{R}_{\theta \kappa \nu }{}^{\alpha } T_{\alpha }{}^{\kappa \theta } T_{\mu } + \tilde{R}_{\theta \kappa \mu }{}^{\alpha } T_{\alpha }{}^{\kappa \theta } T_{\nu }\big) \nonumber\\
&+ 
\bar{h}_{11}^{} \big(\tilde{R}^{\theta \tau }{}_{\nu }{}^{\alpha } T_{\alpha \mu }{}^{\kappa } T_{\tau \kappa \theta } + \tilde{R}_{\nu }{}^{\alpha \theta \tau } T_{\alpha \mu }{}^{\kappa } T_{\tau \kappa \theta } + \tilde{R}^{\theta \tau }{}_{\mu }{}^{\alpha } T_{\alpha \nu }{}^{\kappa } T_{\tau \kappa \theta } \nonumber\\
& -  \tilde{R}^{\alpha }{}_{\kappa }{}^{\tau }{}_{\theta } T_{\alpha \mu }{}^{\kappa } T_{\tau \nu }{}^{\theta } -  \tilde{R}^{\tau }{}_{\theta }{}^{\alpha }{}_{\kappa } T_{\alpha \mu }{}^{\kappa } T_{\tau \nu }{}^{\theta } + \tilde{R}^{\theta \tau }{}_{\mu \kappa } T_{\tau \alpha \theta } T_{\nu }{}^{\alpha \kappa }\big) \nonumber\\
&+ 
2 \bar{h}_{12}^{} \big(\tilde{R}^{\alpha \tau }{}_{\nu }{}^{\theta } T_{\alpha \mu }{}^{\kappa } T_{\tau \kappa \theta } + \tilde{R}^{\alpha \tau }{}_{\mu }{}^{\theta } T_{\alpha \nu }{}^{\kappa } T_{\tau \kappa \theta } -  \tilde{R}^{\alpha \tau }{}_{\kappa \theta } T_{\alpha \mu }{}^{\kappa } T_{\tau \nu }{}^{\theta }\big) \nonumber\\
&+ 
\bar{h}_{13}^{} \big(\tilde{R}^{\alpha \theta }{}_{\nu }{}^{\tau } T_{\alpha \mu }{}^{\kappa } T_{\tau \kappa \theta } + \tilde{R}_{\nu }{}^{\tau \alpha \theta } T_{\alpha \mu }{}^{\kappa } T_{\tau \kappa \theta } + \tilde{R}^{\alpha \theta }{}_{\mu }{}^{\tau } T_{\alpha \nu }{}^{\kappa } T_{\tau \kappa \theta } -  \tilde{R}^{\alpha }{}_{\theta \kappa }{}^{\tau } T_{\alpha \mu }{}^{\kappa } T_{\tau \nu }{}^{\theta } \nonumber\\
& -  \tilde{R}_{\kappa }{}^{\tau \alpha }{}_{\theta } T_{\alpha \mu }{}^{\kappa } T_{\tau \nu }{}^{\theta } -  \tilde{R}_{\kappa }{}^{\tau }{}_{\mu }{}^{\theta } T_{\tau \alpha \theta } T_{\nu }{}^{\alpha \kappa }\big) \nonumber\\
&+ 
\bar{h}_{14}^{} \big[2 \tilde{R}_{\nu }{}^{\theta \alpha \tau } T_{\alpha \mu }{}^{\kappa } T_{\tau \kappa \theta } - 2 \big(\tilde{R}_{\kappa \theta }{}^{\alpha \tau } T_{\alpha \mu }{}^{\kappa } T_{\tau \nu }{}^{\theta } + \tilde{R}_{\kappa }{}^{\theta }{}_{\mu }{}^{\tau } T_{\tau \alpha \theta } T_{\nu }{}^{\alpha \kappa }\big)\big] \nonumber\\
&- 
\bar{h}_{15}^{} \big[\tilde{R}_{\nu }{}^{\tau }{}_{\mu }{}^{\alpha } T_{\alpha }{}^{\kappa \theta } T_{\tau \kappa \theta } +  \tilde{R}^{\alpha }{}_{\mu } T_{\alpha \kappa \theta } T_{\nu }{}^{\kappa \theta }+2 \tilde{R}^{\alpha \kappa } \big(T_{\alpha \nu }{}^{\theta } T_{\kappa \mu \theta } + T_{\alpha \mu }{}^{\theta } T_{\kappa \nu \theta }\big)\big]\nonumber\\
&+ 
\bar{h}_{16}^{} \big(\tilde{R}_{\nu }{}^{\tau }{}_{\mu }{}^{\alpha } T_{\alpha }{}^{\kappa }{}_{\theta } T_{\kappa }{}^{\theta }{}_{\tau } -  \tilde{R}_{\nu }{}^{\alpha } T_{\alpha }{}^{\kappa \theta } T_{\kappa \mu \theta } -  \tilde{R}^{\alpha \kappa } T_{\kappa \nu }{}^{\theta } T_{\theta \mu \alpha } -  \tilde{R}^{\alpha \kappa } T_{\kappa \mu }{}^{\theta } T_{\theta \nu \alpha } -  \tilde{R}^{\alpha \kappa } T_{\kappa \mu \theta } T_{\nu \alpha }{}^{\theta } -  \tilde{R}^{\alpha }{}_{\mu } T_{\kappa \alpha \theta } T_{\nu }{}^{\kappa \theta }\big) \nonumber\\
&+ 
\bar{h}_{17}^{} \big(\tilde{R}_{\nu }{}^{\alpha }{}_{\mu }{}^{\tau } T_{\alpha }{}^{\kappa }{}_{\theta } T_{\kappa }{}^{\theta }{}_{\tau } -  \tilde{R}^{\alpha }{}_{\nu } T_{\alpha }{}^{\kappa \theta } T_{\kappa \mu \theta } -  \tilde{R}^{\alpha }{}_{\mu } T_{\alpha }{}^{\kappa \theta } T_{\kappa \nu \theta } -  \tilde{R}^{\alpha \kappa } T_{\alpha \nu }{}^{\theta } T_{\theta \mu \kappa } -  \tilde{R}^{\alpha \kappa } T_{\alpha \mu }{}^{\theta } T_{\theta \nu \kappa } -  \tilde{R}^{\alpha \kappa } T_{\alpha \mu \theta } T_{\nu \kappa }{}^{\theta }\big) \nonumber\\
&- 
\bar{h}_{18}^{} \big( \tilde{R}_{\nu }{}^{\theta }{}_{\mu \tau } T_{\alpha }{}^{\kappa \tau } T^{\alpha }{}_{\kappa \theta } +  \tilde{R}^{\alpha }{}_{\nu } T_{\kappa \mu }{}^{\theta } T^{\kappa }{}_{\alpha \theta } +  \tilde{R}_{\nu }{}^{\alpha } T_{\kappa \mu }{}^{\theta } T^{\kappa }{}_{\alpha \theta } +  \tilde{R}^{\alpha }{}_{\mu } T_{\kappa \nu }{}^{\theta } T^{\kappa }{}_{\alpha \theta } +  \tilde{R}^{\alpha \kappa } T_{\theta \mu \kappa } T^{\theta }{}_{\nu \alpha } +  \tilde{R}^{\alpha \kappa } T_{\theta \mu \alpha } T^{\theta }{}_{\nu \kappa }\big) \nonumber\\
&+ 
\bar{h}_{19}^{} \big(\tilde{R}^{\alpha \kappa } T_{\theta \mu \kappa } T_{\nu \alpha }{}^{\theta } + \tilde{R}^{\alpha \kappa } T_{\theta \mu \alpha } T_{\nu \kappa }{}^{\theta }- \tilde{R}_{\nu }{}^{\tau }{}_{\mu \theta } T_{\alpha }{}^{\kappa \theta } T_{\kappa }{}^{\alpha }{}_{\tau } -  \tilde{R}^{\alpha }{}_{\nu } T_{\kappa \mu }{}^{\theta } T_{\theta \alpha }{}^{\kappa } -  \tilde{R}_{\nu }{}^{\alpha } T_{\kappa \mu }{}^{\theta } T_{\theta \alpha }{}^{\kappa } -  \tilde{R}^{\alpha }{}_{\mu } T_{\kappa \nu }{}^{\theta } T_{\theta \alpha }{}^{\kappa }\big) \nonumber\\
&+ 
\bar{h}_{20}^{} \big[\tilde{R}^{\alpha \kappa } \big(T_{\nu \mu \kappa } T_{\alpha } + T_{\nu \mu \alpha } T_{\kappa }\big) -  \tilde{R}_{\nu \kappa \mu \alpha } T^{\alpha } T^{\kappa } -  \tilde{R}^{\alpha }{}_{\nu } T_{\alpha } T_{\mu } -  \tilde{R}_{\nu }{}^{\alpha } T_{\alpha } T_{\mu } -  \tilde{R}^{\alpha }{}_{\mu } T_{\alpha } T_{\nu }\big] \nonumber\\
&+ 
\bar{h}_{21}^{} \big[\tilde{R}_{\nu }{}^{\kappa }{}_{\mu \theta } T_{\kappa }{}^{\alpha \theta } T_{\alpha }+ \tilde{R}^{\alpha \kappa } \big(T_{\alpha \kappa \theta } T_{\nu \mu }{}^{\theta } + T_{\alpha \nu \kappa } T_{\mu } + T_{\alpha \mu \kappa } T_{\nu }\big)-  \big(\tilde{R}^{\alpha }{}_{\nu } T_{\alpha \mu }{}^{\kappa } + \tilde{R}^{\alpha }{}_{\mu } T_{\alpha \nu }{}^{\kappa }\big) T_{\kappa }\big]\nonumber\\
& + 
\bar{h}_{22}^{} \big[\tilde{R}_{\nu \theta \mu }{}^{\kappa } T_{\kappa }{}^{\alpha \theta } T_{\alpha } + \tilde{R}^{\alpha \kappa } \big(T_{\kappa \alpha \theta } T_{\nu \mu }{}^{\theta } + T_{\kappa \nu \alpha } T_{\mu } + T_{\kappa \mu \alpha } T_{\nu }\big)-  \big(\tilde{R}_{\nu }{}^{\alpha } T_{\alpha \mu }{}^{\kappa } + \tilde{R}^{\alpha }{}_{\mu } T_{\nu \alpha }{}^{\kappa }\big) T_{\kappa }\big] \nonumber\\
&+ 
\bar{h}_{23}^{} \big[\tilde{R}_{\nu \theta \mu \kappa } T_{\alpha }{}^{\kappa \theta } T^{\alpha } + \big(\tilde{R}^{\alpha }{}_{\nu } T_{\kappa \mu \alpha } -  \tilde{R}_{\nu }{}^{\alpha } T_{\kappa \mu \alpha } + \tilde{R}^{\alpha }{}_{\mu } T_{\kappa \nu \alpha }\big) T^{\kappa } + \tilde{R}^{\alpha \kappa } \big(T_{\theta \alpha \kappa } T_{\nu \mu }{}^{\theta } -  T_{\nu \alpha \kappa } T_{\mu }\big)\big] \nonumber\\
&
-2 \bar{h}_{24}^{} \big(\tilde{R}_{\nu \mu } T_{\alpha }{}^{\kappa \theta } T^{\alpha }{}_{\kappa \theta } + 2 \tilde{R} T_{\alpha \mu }{}^{\kappa } T^{\alpha }{}_{\nu \kappa }\big) 
-2 \bar{h}_{25}^{} \big(\tilde{R}_{\nu \mu } T_{\alpha }{}^{\kappa \theta } T_{\kappa }{}^{\alpha }{}_{\theta } + \tilde{R} T_{\alpha \mu }{}^{\kappa } T_{\kappa \nu }{}^{\alpha } + \tilde{R} T_{\alpha \mu \kappa } T_{\nu }{}^{\alpha \kappa }\big)\nonumber\\
& + 
\bar{h}_{26}^{} \big[2 \tilde{R} T_{\nu \mu }{}^{\alpha } T_{\alpha } - 2 \big(\tilde{R}_{\nu \mu } T_{\alpha } T^{\alpha } + \tilde{R} T_{\mu } T_{\nu }\big)\big]\,.
    \end{align}
Here, the constants $\bar{h}_i$ with $i=1,\ldots,26$ are related to the constants $h_i$ with $i=1,\ldots,26$ appearing in the cubic Lagrangian~\eqref{cubicLagIrr} as follows:
 \begin{eqnarray}
        \bar{h}_{1}^{} &=& \frac{1}{3}\left(h_{5}{}+6 h_{20}{}-h_{7}{}\right)\,,\quad \bar{h}_{2}^{} = \frac{1}{3}\left(h_{6}{}+6 h_{24}{}-h_{8}{}-12 h_{20}{}\right)\,,\quad \bar{h}_{3}^{} =\frac{1}{3}\left(2h_{5}{}-6h_{21}{}\right)\,,\\
        \bar{h}_{4}^{} &=& \frac{1}{3}\left(2 h_{6}{}+2h_{9}{}+12h_{20}{} + 12 h_{22}{}-h_{8}{}- 6 h_{23}{} - 6 h_{24}{}\right)\,,\quad \bar{h}_{5}^{} = 2 h_{24}{}-4 h_{20}{} - 4 h_{21}{} -  \frac{2}{3} h_{6}{} -  \frac{1}{3} h_{8}{}\,,\\
        \bar{h}_{6}^{} &=& \frac{2}{3}\left(h_{5}{}-6h_{20}{} - 3h_{22}{}\right)\,,\quad \bar{h}_{7}^{} =\frac{2}{9} \left(h_{7}{} +  h_{8}{}+ 9 
h_{21}{} + 3 h_{24}{}- 2 h_{5}{} - h_{6}{}-18 h_{13}{} - 3 h_{17}{} - 12 h_{20}{}\right)\,,\\
\bar{h}_{8}^{} &=& \frac{2}{9} (36 h_{13}{} - 3 h_{17}{} + 24 h_{20}{} - 18 
h_{21}{} - 6 h_{24}{} - 2 h_{5}{} -  h_{6}{} + h_{7}{} + h_{8}{})\,,\\
\bar{h}_{9}^{} &=& \frac{1}{9}\left(4 h_{5}{} + 2 
h_{6}{}-72 h_{13}{} - 36 h_{14}{} - 3 h_{16}{} - 24 
h_{20}{} + 36 h_{22}{} - 12 h_{23}{} + 12 h_{24}{} - 2 h_{7}{} - 2 h_{9}{}\right)\,,\\
\bar{h}_{10}^{} &=& \frac{1}{9}\left(2 h_{7}{} + 2 h_{9}{}-36 h_{13}{} - 18 h_{14}{} + 3 h_{16}{} - 
12 h_{20}{} + 18 h_{22}{} - 6 h_{23}{} + 6 h_{24}{} - 4 h_{5}{} - 2 
h_{6}{}\right)\,,\\
\bar{h}_{11}^{} &=& 4 h_{21}{} + 4 h_{22}{} -  \frac{2}{3} h_{7}{}\,,\quad \bar{h}_{12}^{} = \frac{1}{3}\left(h_{9}{}+ 6 h_{23}{}-h_{8}{}-12 h_{22}{}\right)\,,\\
 \bar{h}_{13}^{} &=& \frac{1}{3} (12 h_{21}{} + 12 h_{22}{} - 6 h_{23}{} + 
h_{8}{} + 2 h_{9}{})\,,\quad \bar{h}_{14}^{} = -\,4h_{21}{}\,,\\
\bar{h}_{15}^{} &=& \frac{1}{9} (18 h_{3}{} + 2 h_{5}{} + h_{6}{} + 2 h_{7}{} 
+ h_{8}{} + h_{9}{}-2 h_{10}{} + h_{11}{} - 6 h_{23}{} - 6 
h_{24}{} + 6 h_{25}{})\,,\\
\bar{h}_{16}^{} &=& \frac{1}{9} \bigl[h_{11}{}-2 h_{10}{}+ 2 (6 h_{23}{}
+ 6 h_{24}{}- 18 h_{3}{} - 2 h_{5}{} -  
h_{6}{} - 2 h_{7}{} -  h_{8}{} -  h_{9}{}- 6 h_{25}{} - 18 h_{26}{})\bigr]\,,\\
\bar{h}_{17}^{} &=& \frac{1}{9}\left(h_{11}{}-2 h_{10}{} + 12 h_{23}{} + 12 
h_{24}{} + 6 h_{25}{} + 36 h_{26}{} - 36 h_{3}{} - 4 h_{5}{} - 2 
h_{6}{} - 4 h_{7}{} - 2 h_{8}{} - 2 h_{9}{}\right)\,,\\
\bar{h}_{18}^{} &=& \frac{1}{9} \bigl[5 h_{11}{}- h_{10}{} + 2\bigl(18 h_{3}{} + 2 h_{5}{} + h_{6}{} 
+ 2 h_{7}{} + h_{8}{} + h_{9}{}-6 
h_{23}{} - 6 h_{24}{} - 3 h_{25}{}\bigr)\bigr]\,,\\
\bar{h}_{19}^{} &=& \frac{1}{9} (h_{10}{} + 4 h_{11}{} + 12 h_{23}{} + 12 
h_{24}{} + 6 h_{25}{} - 36 h_{3}{} - 4 h_{5}{} - 2 h_{6}{} - 4 
h_{7}{} - 2 h_{8}{} - 2 h_{9}{})\,,\\
\bar{h}_{20}^{} &=& \frac{1}{9} (9 h_{1}{} + h_{10}{} - 4 h_{11}{} - 3 
h_{16}{} + 6 h_{17}{} - 3 h_{19}{} + 4 h_{5}{} + 2 h_{6}{} - 2 
h_{7}{} -  h_{8}{} - 2 h_{9}{})\,,\\
\bar{h}_{21}^{} &=& \frac{1}{9}\left(3 h_{11}{} + 18 h_{15}{}-3 h_{10}{} - 3 
h_{18}{} + 6 h_{19}{} + 12 h_{21}{} - 12 h_{22}{} + 6 h_{23}{} + 6 
h_{25}{} + 24 h_{26}{} -  h_{8}{} + 4 h_{9}{}\right)\,,\\
\bar{h}_{22}^{} &=& \frac{1}{9}\left(3 h_{11}{}-3 h_{10}{} - 18 h_{15}{} + 3 
h_{18}{} + 3 h_{19}{} - 12 h_{21}{} + 12 h_{22}{} - 6 h_{23}{} - 6 
h_{25}{} - 24 h_{26}{} + h_{8}{} + 2 h_{9}{}\right)\,,\\
\bar{h}_{23}^{} &=& \frac{1}{9} (18 h_{15}{} + 6 h_{18}{} - 3 h_{19}{} + 12 
h_{21}{} - 12 h_{22}{} + 6 h_{23}{} + 6 h_{25}{} + 24 h_{26}{} + 2 
h_{8}{} - 2 h_{9}{})\,,\\
\bar{h}_{24}^{} &=& \frac{1}{18} (12 h_{12}{} + 6 h_{23}{} + 6 h_{24}{} - 36
h_{3}{} - 36 h_{4}{} - 2 h_{5}{} -  h_{6}{} - 2 h_{7}{} -  h_{8}{} -  
h_{9}{})\,,\\
\bar{h}_{25}^{} &=& \frac{1}{9}\left(36 
h_{3}{} + 36 h_{4}{} + 2 h_{5}{} + h_{6}{} + 2 h_{7}{} + h_{8}{} + 
h_{9}{}+6 h_{12}{}- 6 h_{23}{} - 6 h_{24}{}\right)\,,\\
\bar{h}_{26}^{} &=& \frac{1}{9}\left(9 h_{2}{} + h_{9}{}+ h_{11}{}+ 3 
h_{19}{}- h_{10}{} - 6 h_{12}{}\right)\,.
    \end{eqnarray}
\normalsize

\section{\texorpdfstring{$\Lambda$}{Lambda}CDM Model} \label{sec:lcdm}

\begin{figure}[htbp]
    \centering
    \includegraphics[scale = 0.5]{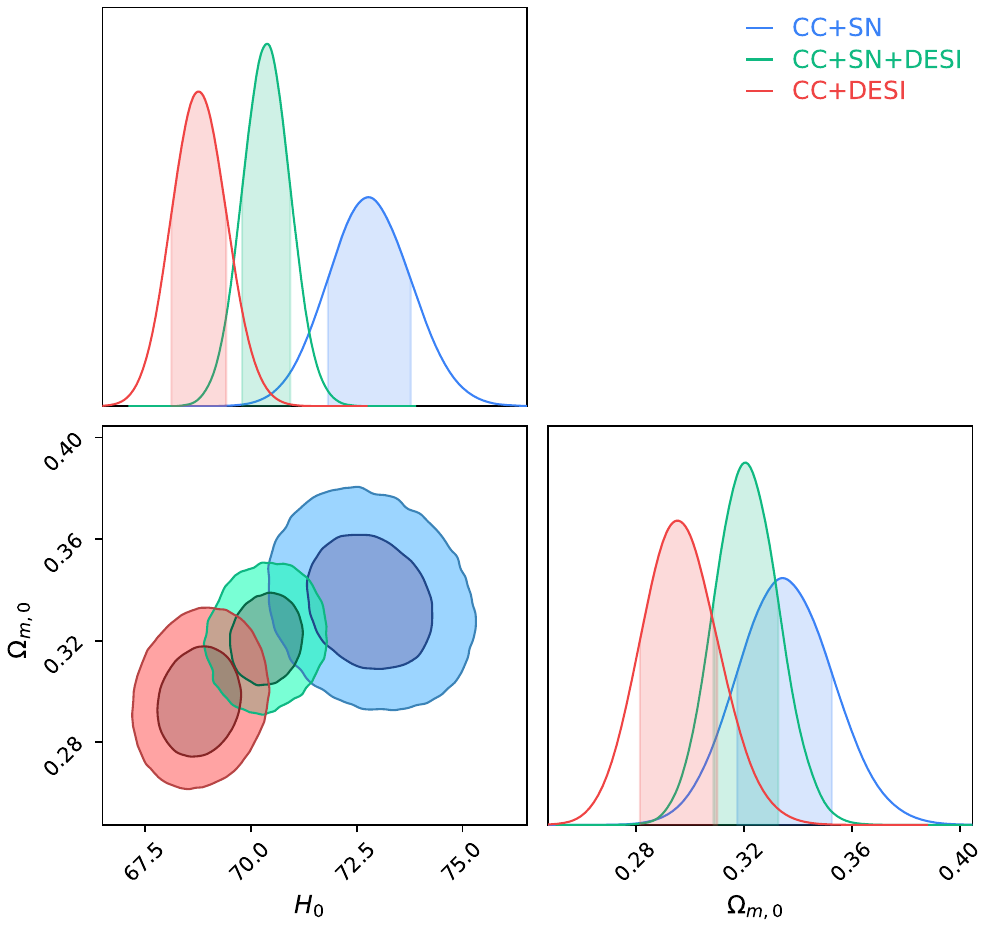}
    \caption{Confidence contours and posterior distributions for the $\Lambda$CDM model. The blue contours represent the CC+SN dataset, the red contours illustrate CC+DESI, while the green contours correspond to the combination CC+SN+DESI.}
\label{fig:lcdm}

\end{figure}

This section presents the results for the standard $\Lambda$CDM model. The posterior distributions for $H_0$ and $\Omega_{m,0}$ are shown in Fig.~\ref{fig:lcdm}, while the corresponding numerical values, including the nuisance parameter $M$, are reported in Table~\ref{tab:lcdm_otputs}. The primary purpose of this section is to establish a benchmark against which the extended models can be compared. To this end, the values of the information criteria—AIC and BIC—are listed in Table~\ref{tab:lcdm_outputs2}. These serve as reference values for computing the $\Delta$AIC and $\Delta$BIC, which quantify the relative performance of the extended models.

\begin{table}[h]
    \centering
    \label{tab:lcdm_otputs}
    \begin{tabular}{cccccc}
        \hline
        \rule{0pt}{3ex}
		Model & $H_0\, [{\rm km\, s}^{-1} {\rm Mpc}^{-1}]$ & $\Omega_{m,0}$ & $M$ \\ [4 pt]
		\hline
        \rule{0pt}{4ex}
		CC+SN & $72.80^{+0.97}_{-0.98}$ & $0.334^{+0.019}_{-0.016}$ & $-19.262^{+0.030}_{-0.027}$\\ 
		CC+SN+DESI & $70.41^{+0.52}_{-0.62}$ & $0.321\pm 0.012$ & $-19.340^{+0.020}_{-0.017}$\\ 
		CC+DESI & $68.75^{+0.66}_{-0.63}$ & $0.295^{+0.015}_{-0.014}$& -- \\[4 pt] 

		\hline

    \end{tabular}
    \caption{Exact results for the $\Lambda$CDM model, including the parameters $H_0$ and $\Omega_{m,0}$ for different dataset combinations.}
\end{table}

\begin{table}[h]
\centering
 
    \label{tab:lcdm_outputs2}
\begin{tabular}{l c c  c}
\hline
\rule{0pt}{3 ex}
Data sets & $\chi^2_{\textrm{min}}$ & AIC & BIC \\ 
		\hline
        \rule{0pt}{4ex}
CC+SN & 1539.22 & 1545.22 & 1548.92 \\
CC+SN+DESI & 1567.30 & 1573.30 & 1577.01 \\
CC+DESI & 19.08 &25.08 & 23.37 \\
\hline
\end{tabular}
\caption{Minimum $\chi^2_{min}$ values along with AIC and BIC statistics for the $\Lambda$CDM model, used as reference benchmarks for model comparison.}
\end{table}

\bibliographystyle{utphys}
\bibliography{references}

\providecommand{\href}[2]{#2}\begingroup\raggedright\begin{thebibliography}{10}

\bibitem{Misner:1974qy}
C.~W. Misner, K.~S. Thorne, and J.~A. Wheeler, {\em {Gravitation}}.
\newblock W. H. Freeman, San Francisco, 1973.

\bibitem{Clifton:2011jh}
T.~Clifton, P.~G. Ferreira, A.~Padilla, and C.~Skordis, ``{Modified Gravity and Cosmology},'' \href{http://dx.doi.org/10.1016/j.physrep.2012.01.001}{{\em Phys. Rept.} {\bf 513} (2012)  1--189},
\href{http://arxiv.org/abs/1106.2476}{{\tt arXiv:1106.2476 [astro-ph.CO]}}.

\bibitem{Peebles:2002gy}
P.~J.~E. Peebles and B.~Ratra, ``{The Cosmological constant and dark energy},'' \href{http://dx.doi.org/10.1103/RevModPhys.75.559}{{\em Rev. Mod. Phys.} {\bf 75} (2003)  559--606}, \href{http://arxiv.org/abs/astro-ph/0207347}{{\tt arXiv:astro-ph/0207347 [astro-ph]}}.
[,592(2002)].

\bibitem{Baudis:2016qwx}
L.~Baudis, ``{Dark matter detection},''
\href{http://dx.doi.org/10.1088/0954-3899/43/4/044001}{{\em J. Phys.} {\bf G43} (2016) no.~4, 044001}.

\bibitem{Bertone:2004pz}
G.~Bertone, D.~Hooper, and J.~Silk, ``{Particle dark matter: Evidence, candidates and constraints},'' \href{http://dx.doi.org/10.1016/j.physrep.2004.08.031}{{\em Phys. Rept.} {\bf 405} (2005)  279--390},
\href{http://arxiv.org/abs/hep-ph/0404175}{{\tt arXiv:hep-ph/0404175 [hep-ph]}}.

\bibitem{Copeland:2006wr}
E.~J. Copeland, M.~Sami, and S.~Tsujikawa, ``{Dynamics of dark energy},'' \href{http://dx.doi.org/10.1142/S021827180600942X}{{\em Int. J. Mod. Phys.} {\bf D15} (2006)  1753--1936},
\href{http://arxiv.org/abs/hep-th/0603057}{{\tt arXiv:hep-th/0603057 [hep-th]}}.

\bibitem{Weinberg:1988cp}
S.~Weinberg, ``{The Cosmological Constant Problem},'' \href{http://dx.doi.org/10.1103/RevModPhys.61.1}{{\em Rev. Mod. Phys.} {\bf 61} (1989)  1--23}.

\bibitem{Gaitskell:2004gd}
R.~J. Gaitskell, ``{Direct detection of dark matter},'' \href{http://dx.doi.org/10.1146/annurev.nucl.54.070103.181244}{{\em Ann. Rev. Nucl. Part. Sci.} {\bf 54} (2004)  315--359}.

\bibitem{DiBari:2013dna}
P.~Di~Bari, S.~F. King, and A.~Merle, ``{Dark Radiation or Warm Dark Matter from long lived particle decays in the light of Planck},'' \href{http://dx.doi.org/10.1016/j.physletb.2013.06.003}{{\em Phys. Lett. B} {\bf 724} (2013)  77--83}, \href{http://arxiv.org/abs/1303.6267}{{\tt arXiv:1303.6267 [hep-ph]}}.

\bibitem{doi:10.1073/pnas.90.11.4766}
C.~L. Bennett, N.~W. Boggess, E.~S. Cheng, M.~G. Hauser, T.~Kelsall, J.~C. Mather, S.~H. Moseley, T.~L. Murdock, R.~A. Shafer, R.~F. Silverberg, G.~F. Smoot, R.~Weiss, and E.~L. Wright, ``Scientific results from the cosmic background explorer (cobe),'' \href{http://dx.doi.org/10.1073/pnas.90.11.4766}{{\em Proceedings of the National Academy of Sciences} {\bf 90} (1993) no.~11, 4766--4773}, \href{http://arxiv.org/abs/https://www.pnas.org/doi/pdf/10.1073/pnas.90.11.4766}{{\tt https://www.pnas.org/doi/pdf/10.1073/pnas.90.11.4766}}. \url{https://www.pnas.org/doi/abs/10.1073/pnas.90.11.4766}.

\bibitem{2020ApJS..247...69E}
P.~R.~M. {Eisenhardt}, F.~{Marocco}, J.~W. {Fowler}, A.~M. {Meisner}, J.~D. {Kirkpatrick}, N.~{Garcia}, T.~H. {Jarrett}, R.~{Koontz}, E.~J. {Marchese}, S.~A. {Stanford}, D.~{Caselden}, M.~C. {Cushing}, R.~M. {Cutri}, J.~K. {Faherty}, C.~R. {Gelino}, A.~H. {Gonzalez}, A.~{Mainzer}, B.~{Mobasher}, D.~J. {Schlegel}, D.~{Stern}, H.~I. {Teplitz}, and E.~L. {Wright}, ``{The CatWISE Preliminary Catalog: Motions from WISE and NEOWISE Data},'' \href{http://dx.doi.org/10.3847/1538-4365/ab7f2a}{{\em ApJL} {\bf 247} (2020) no.~2, 69}, \href{http://arxiv.org/abs/1908.08902}{{\tt arXiv:1908.08902 [astro-ph.IM]}}.

\bibitem{2007ApJ...660L..81E}
H.~K. {Eriksen}, A.~J. {Banday}, K.~M. {G{\'o}rski}, F.~K. {Hansen}, and P.~B. {Lilje}, ``{Hemispherical Power Asymmetry in the Third-Year Wilkinson Microwave Anisotropy Probe Sky Maps},'' \href{http://dx.doi.org/10.1086/518091}{{\em ApJL} {\bf 660} (2007) no.~2, L81--L84}, \href{http://arxiv.org/abs/astro-ph/0701089}{{\tt arXiv:astro-ph/0701089 [astro-ph]}}.

\bibitem{Aghanim:2018eyx}
{\bf Planck} Collaboration, N.~Aghanim {\em et al.}, ``{Planck 2018 results. VI. Cosmological parameters},''
\href{http://arxiv.org/abs/1807.06209}{{\tt arXiv:1807.06209 [astro-ph.CO]}}.

\bibitem{Perivolaropoulos:2021jda}
L.~Perivolaropoulos and F.~Skara, ``{Challenges for \ensuremath{\Lambda}CDM: An update},'' \href{http://dx.doi.org/10.1016/j.newar.2022.101659}{{\em New Astron. Rev.} {\bf 95} (2022)  101659}, \href{http://arxiv.org/abs/2105.05208}{{\tt arXiv:2105.05208 [astro-ph.CO]}}.

\bibitem{CosmoVerse:2025txj}
{\bf CosmoVerse} Collaboration, E.~Di~Valentino {\em et al.}, ``{The CosmoVerse White Paper: Addressing observational tensions in cosmology with systematics and fundamental physics},'' \href{http://arxiv.org/abs/2504.01669}{{\tt arXiv:2504.01669 [astro-ph.CO]}}.

\bibitem{DiValentino:2020vhf}
E.~Di~Valentino {\em et al.}, ``{Snowmass2021 - Letter of interest cosmology intertwined I: Perspectives for the next decade},'' \href{http://dx.doi.org/10.1016/j.astropartphys.2021.102606}{{\em Astropart. Phys.} {\bf 131} (2021)  102606}, \href{http://arxiv.org/abs/2008.11283}{{\tt arXiv:2008.11283 [astro-ph.CO]}}.

\bibitem{DiValentino:2020zio}
E.~Di~Valentino {\em et al.}, ``{Snowmass2021 - Letter of interest cosmology intertwined II: The hubble constant tension},'' \href{http://dx.doi.org/10.1016/j.astropartphys.2021.102605}{{\em Astropart. Phys.} {\bf 131} (2021)  102605}, \href{http://arxiv.org/abs/2008.11284}{{\tt arXiv:2008.11284 [astro-ph.CO]}}.

\bibitem{DiValentino:2020vvd}
E.~Di~Valentino {\em et al.}, ``{Cosmology Intertwined III: $f \sigma_8$ and $S_8$},'' \href{http://dx.doi.org/10.1016/j.astropartphys.2021.102604}{{\em Astropart. Phys.} {\bf 131} (2021)  102604}, \href{http://arxiv.org/abs/2008.11285}{{\tt arXiv:2008.11285 [astro-ph.CO]}}.

\bibitem{DiValentino:2020srs}
E.~Di~Valentino {\em et al.}, ``{Snowmass2021 - Letter of interest cosmology intertwined IV: The age of the universe and its curvature},'' \href{http://dx.doi.org/10.1016/j.astropartphys.2021.102607}{{\em Astropart. Phys.} {\bf 131} (2021)  102607}, \href{http://arxiv.org/abs/2008.11286}{{\tt arXiv:2008.11286 [astro-ph.CO]}}.

\bibitem{Riess:2019cxk}
A.~G. Riess, S.~Casertano, W.~Yuan, L.~M. Macri, and D.~Scolnic, ``{Large Magellanic Cloud Cepheid Standards Provide a 1\% Foundation for the Determination of the Hubble Constant and Stronger Evidence for Physics beyond $\Lambda$CDM},'' \href{http://dx.doi.org/10.3847/1538-4357/ab1422}{{\em Astrophys. J.} {\bf 876} (2019) no.~1, 85}, \href{http://arxiv.org/abs/1903.07603}{{\tt arXiv:1903.07603 [astro-ph.CO]}}.

\bibitem{Anderson:2023aga}
R.~I. Anderson, N.~W. Koblischke, and L.~Eyer, ``{Reconciling astronomical distance scales with variable red giant stars},'' \href{http://arxiv.org/abs/2303.04790}{{\tt arXiv:2303.04790 [astro-ph.CO]}}.

\bibitem{Wong:2019kwg}
K.~C. Wong {\em et al.}, ``{H0LiCOW \textendash{} XIII. A 2.4 per cent measurement of H0 from lensed quasars: 5.3\ensuremath{\sigma} tension between early- and late-Universe probes},'' \href{http://dx.doi.org/10.1093/mnras/stz3094}{{\em Mon. Not. Roy. Astron. Soc.} {\bf 498} (2020) no.~1, 1420--1439}, \href{http://arxiv.org/abs/1907.04869}{{\tt arXiv:1907.04869 [astro-ph.CO]}}.

\bibitem{ACT:2023kun}
{\bf ACT} Collaboration, M.~S. Madhavacheril {\em et al.}, ``{The Atacama Cosmology Telescope: DR6 Gravitational Lensing Map and Cosmological Parameters},'' \href{http://arxiv.org/abs/2304.05203}{{\tt arXiv:2304.05203 [astro-ph.CO]}}.

\bibitem{Schoneberg:2022ggi}
N.~Sch\"oneberg, L.~Verde, H.~Gil-Mar\'\i{}n, and S.~Brieden, ``{BAO+BBN revisited \textemdash{} growing the Hubble tension with a 0.7 km/s/Mpc constraint},'' \href{http://dx.doi.org/10.1088/1475-7516/2022/11/039}{{\em JCAP} {\bf 11} (2022)  039}, \href{http://arxiv.org/abs/2209.14330}{{\tt arXiv:2209.14330 [astro-ph.CO]}}.

\bibitem{Riess:2020sih}
A.~G. Riess, ``{The Expansion of the Universe is Faster than Expected},'' \href{http://dx.doi.org/10.1038/s42254-019-0137-0}{{\em Nature Rev. Phys.} {\bf 2} (2019) no.~1, 10--12}, \href{http://arxiv.org/abs/2001.03624}{{\tt arXiv:2001.03624 [astro-ph.CO]}}.

\bibitem{Pesce:2020xfe}
D.~W. Pesce {\em et al.}, ``{The Megamaser Cosmology Project. XIII. Combined Hubble constant constraints},'' \href{http://dx.doi.org/10.3847/2041-8213/ab75f0}{{\em Astrophys. J. Lett.} {\bf 891} (2020) no.~1, L1}, \href{http://arxiv.org/abs/2001.09213}{{\tt arXiv:2001.09213 [astro-ph.CO]}}.

\bibitem{deJaeger:2020zpb}
T.~de~Jaeger, B.~E. Stahl, W.~Zheng, A.~V. Filippenko, A.~G. Riess, and L.~Galbany, ``{A measurement of the Hubble constant from Type II supernovae},'' \href{http://dx.doi.org/10.1093/mnras/staa1801}{{\em Mon. Not. Roy. Astron. Soc.} {\bf 496} (2020) no.~3, 3402--3411}, \href{http://arxiv.org/abs/2006.03412}{{\tt arXiv:2006.03412 [astro-ph.CO]}}.

\bibitem{Abdalla:2022yfr}
E.~Abdalla {\em et al.}, ``{Cosmology intertwined: A review of the particle physics, astrophysics, and cosmology associated with the cosmological tensions and anomalies},'' \href{http://dx.doi.org/10.1016/j.jheap.2022.04.002}{{\em JHEAp} {\bf 34} (2022)  49--211}, \href{http://arxiv.org/abs/2203.06142}{{\tt arXiv:2203.06142 [astro-ph.CO]}}.

\bibitem{Bernal:2016gxb}
J.~L. Bernal, L.~Verde, and A.~G. Riess, ``{The trouble with $H_0$},'' \href{http://dx.doi.org/10.1088/1475-7516/2016/10/019}{{\em JCAP} {\bf 10} (2016)  019}, \href{http://arxiv.org/abs/1607.05617}{{\tt arXiv:1607.05617 [astro-ph.CO]}}.

\bibitem{Sotiriou:2008rp}
T.~P. Sotiriou and V.~Faraoni, ``{f(R) Theories Of Gravity},'' \href{http://dx.doi.org/10.1103/RevModPhys.82.451}{{\em Rev. Mod. Phys.} {\bf 82} (2010)  451--497},
\href{http://arxiv.org/abs/0805.1726}{{\tt arXiv:0805.1726 [gr-qc]}}.

\bibitem{Nojiri:2010wj}
S.~Nojiri and S.~D. Odintsov, ``{Unified cosmic history in modified gravity: from F(R) theory to Lorentz non-invariant models},'' \href{http://dx.doi.org/10.1016/j.physrep.2011.04.001}{{\em Phys. Rept.} {\bf 505} (2011)  59--144}, \href{http://arxiv.org/abs/1011.0544}{{\tt arXiv:1011.0544 [gr-qc]}}.

\bibitem{Bamba:2012cp}
K.~Bamba, S.~Capozziello, S.~Nojiri, and S.~D. Odintsov, ``{Dark energy cosmology: the equivalent description via different theoretical models and cosmography tests},'' \href{http://dx.doi.org/10.1007/s10509-012-1181-8}{{\em Astrophys. Space Sci.} {\bf 342} (2012)  155--228}, \href{http://arxiv.org/abs/1205.3421}{{\tt arXiv:1205.3421 [gr-qc]}}.

\bibitem{Nojiri:2017ncd}
S.~Nojiri, S.~D. Odintsov, and V.~K. Oikonomou, ``{Modified Gravity Theories on a Nutshell: Inflation, Bounce and Late-time Evolution},'' \href{http://dx.doi.org/10.1016/j.physrep.2017.06.001}{{\em Phys. Rept.} {\bf 692} (2017)  1--104}, \href{http://arxiv.org/abs/1705.11098}{{\tt arXiv:1705.11098 [gr-qc]}}.

\bibitem{CANTATA:2021ktz}
{\bf CANTATA} Collaboration, E.~N. Saridakis, R.~Lazkoz, V.~Salzano, P.~Vargas~Moniz, S.~Capozziello, J.~Beltr\'an~Jim\'enez, M.~De~Laurentis, and G.~J. Olmo, eds., \href{http://dx.doi.org/10.1007/978-3-030-83715-0}{{\em {Modified Gravity and Cosmology}: {An Update by the CANTATA Network}}}.
\newblock Springer, 2021.
\newblock \href{http://arxiv.org/abs/2105.12582}{{\tt arXiv:2105.12582 [gr-qc]}}.

\bibitem{Krishnan:2020vaf}
C.~Krishnan, E.~O. Colg\'ain, M.~M. Sheikh-Jabbari, and T.~Yang, ``{Running Hubble Tension and a H0 Diagnostic},'' \href{http://dx.doi.org/10.1103/PhysRevD.103.103509}{{\em Phys. Rev. D} {\bf 103} (2021) no.~10, 103509}, \href{http://arxiv.org/abs/2011.02858}{{\tt arXiv:2011.02858 [astro-ph.CO]}}.

\bibitem{Colgain:2022nlb}
E.~O. Colg\'ain, M.~M. Sheikh-Jabbari, R.~Solomon, G.~Bargiacchi, S.~Capozziello, M.~G. Dainotti, and D.~Stojkovic, ``{Revealing intrinsic flat \ensuremath{\Lambda}CDM biases with standardizable candles},'' \href{http://dx.doi.org/10.1103/PhysRevD.106.L041301}{{\em Phys. Rev. D} {\bf 106} (2022) no.~4, L041301}, \href{http://arxiv.org/abs/2203.10558}{{\tt arXiv:2203.10558 [astro-ph.CO]}}.

\bibitem{Malekjani:2023dky}
M.~Malekjani, R.~M. Conville, E.~O. Colg\'ain, S.~Pourojaghi, and M.~M. Sheikh-Jabbari, ``{Negative Dark Energy Density from High Redshift Pantheon+ Supernovae},'' \href{http://arxiv.org/abs/2301.12725}{{\tt arXiv:2301.12725 [astro-ph.CO]}}.

\bibitem{Ren:2022aeo}
X.~Ren, S.-F. Yan, Y.~Zhao, Y.-F. Cai, and E.~N. Saridakis, ``{Gaussian processes and effective field theory of $f(T)$ gravity under the $H_0$ tension},'' \href{http://dx.doi.org/10.3847/1538-4357/ac6ba5}{{\em Astrophys. J.} {\bf 932} (2022)  2}, \href{http://arxiv.org/abs/2203.01926}{{\tt arXiv:2203.01926 [astro-ph.CO]}}.

\bibitem{Dainotti:2021pqg}
M.~G. Dainotti, B.~De~Simone, T.~Schiavone, G.~Montani, E.~Rinaldi, and G.~Lambiase, ``{On the Hubble constant tension in the SNe Ia Pantheon sample},'' \href{http://dx.doi.org/10.3847/1538-4357/abeb73}{{\em Astrophys. J.} {\bf 912} (2021) no.~2, 150}, \href{http://arxiv.org/abs/2103.02117}{{\tt arXiv:2103.02117 [astro-ph.CO]}}.

\bibitem{Addazi:2021xuf}
A.~Addazi {\em et al.}, ``{Quantum gravity phenomenology at the dawn of the multi-messenger era\textemdash{}A review},'' \href{http://dx.doi.org/10.1016/j.ppnp.2022.103948}{{\em Prog. Part. Nucl. Phys.} {\bf 125} (2022)  103948}, \href{http://arxiv.org/abs/2111.05659}{{\tt arXiv:2111.05659 [hep-ph]}}.

\bibitem{Schoneberg:2021qvd}
N.~Sch\"oneberg, G.~Franco~Abell\'an, A.~P\'erez~S\'anchez, S.~J. Witte, V.~Poulin, and J.~Lesgourgues, ``{The H0 Olympics: A fair ranking of proposed models},'' \href{http://dx.doi.org/10.1016/j.physrep.2022.07.001}{{\em Phys. Rept.} {\bf 984} (2022)  1--55}, \href{http://arxiv.org/abs/2107.10291}{{\tt arXiv:2107.10291 [astro-ph.CO]}}.

\bibitem{e24c2fd0-bd80-346c-b6f7-f15690a270c6}
H.~GUTFREUND and J.~RENN, {\em The Formative Years of Relativity: The History and Meaning of Einstein's Princeton Lectures}.
\newblock Princeton University Press, 2017.
\newblock \url{http://www.jstor.org/stable/j.ctt1vxm7ts}.

\bibitem{Bahamonde:2021gfp}
S.~Bahamonde, K.~F. Dialektopoulos, C.~Escamilla-Rivera, G.~Farrugia, V.~Gakis, M.~Hendry, M.~Hohmann, J.~Levi~Said, J.~Mifsud, and E.~Di~Valentino, ``{Teleparallel gravity: from theory to cosmology},'' \href{http://dx.doi.org/10.1088/1361-6633/ac9cef}{{\em Rept. Prog. Phys.} {\bf 86} (2023) no.~2, 026901}, \href{http://arxiv.org/abs/2106.13793}{{\tt arXiv:2106.13793 [gr-qc]}}.

\bibitem{Aldrovandi:2013wha}
R.~Aldrovandi and J.~G. Pereira, \href{http://dx.doi.org/10.1007/978-94-007-5143-9}{{\em {Teleparallel Gravity}}}, vol.~173.
\newblock Springer, Dordrecht,
2013.
\newblock

\bibitem{Cai:2015emx}
Y.-F. Cai, S.~Capozziello, M.~De~Laurentis, and E.~N. Saridakis, ``{f(T) teleparallel gravity and cosmology},'' \href{http://dx.doi.org/10.1088/0034-4885/79/10/106901}{{\em Rept. Prog. Phys.} {\bf 79} (2016) no.~10, 106901},
\href{http://arxiv.org/abs/1511.07586}{{\tt arXiv:1511.07586 [gr-qc]}}.

\bibitem{Krssak:2018ywd}
M.~Krssak, R.~J. Van Den~Hoogen, J.~G. Pereira, C.~G. Boehmer, and A.~A. Coley, ``{Teleparallel Theories of Gravity: Illuminating a Fully Invariant Approach},''
\href{http://arxiv.org/abs/1810.12932}{{\tt arXiv:1810.12932 [gr-qc]}}.

\bibitem{Nester:1998mp}
J.~M. Nester and H.-J. Yo, ``{Symmetric teleparallel general relativity},'' {\em Chin. J. Phys.} {\bf 37} (1999)  113,
\href{http://arxiv.org/abs/gr-qc/9809049}{{\tt arXiv:gr-qc/9809049 [gr-qc]}}.

\bibitem{BeltranJimenez:2019esp}
J.~Beltr\'an~Jim\'enez, L.~Heisenberg, and T.~S. Koivisto, ``{The Geometrical Trinity of Gravity},'' \href{http://dx.doi.org/10.3390/universe5070173}{{\em Universe} {\bf 5} (2019) no.~7, 173}, \href{http://arxiv.org/abs/1903.06830}{{\tt arXiv:1903.06830 [hep-th]}}.

\bibitem{Hehl:1994ue}
F.~W. Hehl, J.~D. McCrea, E.~W. Mielke, and Y.~Ne'eman, ``{Metric-Affine gauge theory of gravity: Field equations, Noether identities, world spinors, and breaking of dilation invariance},'' \href{http://dx.doi.org/10.1016/0370-1573(94)00111-F}{{\em Phys. Rept.} {\bf 258} (1995)  1--171},
\href{http://arxiv.org/abs/gr-qc/9402012}{{\tt arXiv:gr-qc/9402012 [gr-qc]}}.

\bibitem{Blagojevic:2012bc}
M.~Blagojevic and F.~W. Hehl, ``{Gauge Theories of Gravitation},'' \href{http://arxiv.org/abs/1210.3775}{{\tt arXiv:1210.3775 [gr-qc]}}.

\bibitem{BeltranJimenez:2019hrm}
J.~Beltr\'an~Jim\'enez and F.~J. Maldonado~Torralba, ``{Revisiting the stability of quadratic Poincar\'e gauge gravity},'' \href{http://dx.doi.org/10.1140/epjc/s10052-020-8163-8}{{\em Eur. Phys. J. C} {\bf 80} (2020) no.~7, 611}, \href{http://arxiv.org/abs/1910.07506}{{\tt arXiv:1910.07506 [gr-qc]}}.

\bibitem{Bahamonde:2024sqo}
S.~Bahamonde and J.~Gigante~Valcarcel, ``{Stability of Poincar\'e gauge theory with cubic order invariants},'' \href{http://dx.doi.org/10.1103/PhysRevD.109.104075}{{\em Phys. Rev. D} {\bf 109} (2024) no.~10, 104075}, \href{http://arxiv.org/abs/2402.08937}{{\tt arXiv:2402.08937 [gr-qc]}}.

\bibitem{TSAMPARLIS197927}
M.~Tsamparlis, ``Cosmological principle and torsion,'' \href{http://dx.doi.org/https://doi.org/10.1016/0375-9601(79)90265-2}{{\em Physics Letters A} {\bf 75} (1979) no.~1, 27--28}. \url{https://www.sciencedirect.com/science/article/pii/0375960179902652}.

\bibitem{Cembranos:2018ipn}
J.~A.~R. Cembranos, J.~Gigante~Valcarcel, and F.~J. Maldonado~Torralba, ``{Fermion dynamics in torsion theories},'' \href{http://dx.doi.org/10.1088/1475-7516/2019/04/039}{{\em JCAP} {\bf 04} (2019)  039}, \href{http://arxiv.org/abs/1805.09577}{{\tt arXiv:1805.09577 [gr-qc]}}.

\bibitem{Puetzfeld:2014qba}
D.~Puetzfeld and Y.~N. Obukhov, ``{Equations of motion in metric-affine gravity: A covariant unified framework},'' \href{http://dx.doi.org/10.1103/PhysRevD.90.084034}{{\em Phys. Rev. D} {\bf 90} (2014) no.~8, 084034}, \href{http://arxiv.org/abs/1408.5669}{{\tt arXiv:1408.5669 [gr-qc]}}.

\bibitem{Iosifidis:2020gth}
D.~Iosifidis, ``{Cosmological Hyperfluids, Torsion and Non-metricity},'' \href{http://dx.doi.org/10.1140/epjc/s10052-020-08634-z}{{\em Eur. Phys. J. C} {\bf 80} (2020) no.~11, 1042}, \href{http://arxiv.org/abs/2003.07384}{{\tt arXiv:2003.07384 [gr-qc]}}.

\bibitem{Iosifidis:2023but}
D.~Iosifidis, ``{Metric-Affine Cosmologies: kinematics of Perfect (Ideal) Cosmological Hyperfluids and first integrals},'' \href{http://dx.doi.org/10.1088/1475-7516/2023/09/045}{{\em JCAP} {\bf 09} (2023)  045}, \href{http://arxiv.org/abs/2301.09868}{{\tt arXiv:2301.09868 [gr-qc]}}.

\bibitem{Dyer:2024kvo}
T.~Dyer, W.~Barker, and D.~Iosifidis, ``{Complete background cosmology of parity-even quadratic metric-affine gravity},'' \href{http://arxiv.org/abs/2412.15329}{{\tt arXiv:2412.15329 [gr-qc]}}.

\bibitem{Brout:2021mpj}
D.~Brout {\em et al.}, ``{The Pantheon+ Analysis: SuperCal-fragilistic Cross Calibration, Retrained SALT2 Light-curve Model, and Calibration Systematic Uncertainty},'' \href{http://dx.doi.org/10.3847/1538-4357/ac8bcc}{{\em Astrophys. J.} {\bf 938} (2022) no.~2, 111}, \href{http://arxiv.org/abs/2112.03864}{{\tt arXiv:2112.03864 [astro-ph.CO]}}.

\bibitem{Riess:2021jrx}
A.~G. Riess {\em et al.}, ``A comprehensive measurement of the local value of the hubble constant with 1 km s$^{-1}$ mpc$^{-1}$ uncertainty from the hubble space telescope and the sh0es team,'' \href{http://dx.doi.org/10.3847/2041-8213/ac5c5b}{{\em Astrophys. J. Lett.} {\bf 934} (2022) no.~1, L7}, \href{http://arxiv.org/abs/2112.04510}{{\tt arXiv:2112.04510 [astro-ph.CO]}}.

\bibitem{Scolnic:2021amr}
D.~Scolnic {\em et al.}, ``{The Pantheon+ Analysis: The Full Data Set and Light-curve Release},'' \href{http://dx.doi.org/10.3847/1538-4357/ac8b7a}{{\em Astrophys. J.} {\bf 938} (2022) no.~2, 113}, \href{http://arxiv.org/abs/2112.03863}{{\tt arXiv:2112.03863 [astro-ph.CO]}}.

\bibitem{Keeley:2022iba}
R.~E. Keeley, A.~Shafieloo, and B.~L'Huillier, ``{An Analysis of Variance of the Pantheon+ Dataset: Systematics in the Covariance Matrix?},'' \href{http://dx.doi.org/10.3390/universe10120439}{{\em Universe} {\bf 10} (2024) no.~12, 439}, \href{http://arxiv.org/abs/2212.07917}{{\tt arXiv:2212.07917 [astro-ph.CO]}}.

\bibitem{Perivolaropoulos:2023iqj}
L.~Perivolaropoulos and F.~Skara, ``{On the homogeneity of SnIa absolute magnitude in the Pantheon+~sample},'' \href{http://dx.doi.org/10.1093/mnras/stad451}{{\em Mon. Not. Roy. Astron. Soc.} {\bf 520} (2023) no.~4, 5110--5125}, \href{http://arxiv.org/abs/2301.01024}{{\tt arXiv:2301.01024 [astro-ph.CO]}}.

\bibitem{Moresco:2020fbm}
M.~Moresco, R.~Jimenez, L.~Verde, A.~Cimatti, and L.~Pozzetti, ``{Setting the Stage for Cosmic Chronometers. II. Impact of Stellar Population Synthesis Models Systematics and Full Covariance Matrix},'' \href{http://dx.doi.org/10.3847/1538-4357/ab9eb0}{{\em Astrophys. J.} {\bf 898} (2020) no.~1, 82}, \href{http://arxiv.org/abs/2003.07362}{{\tt arXiv:2003.07362 [astro-ph.GA]}}.

\bibitem{Moresco:2023zys}
M.~Moresco, ``{Addressing the Hubble tension with cosmic chronometers},'' \href{http://arxiv.org/abs/2307.09501}{{\tt arXiv:2307.09501 [astro-ph.CO]}}.

\bibitem{Moresco:2022phi}
M.~Moresco {\em et al.}, ``{Unveiling the Universe with emerging cosmological probes},'' \href{http://dx.doi.org/10.1007/s41114-022-00040-z}{{\em Living Rev. Rel.} {\bf 25} (2022) no.~1, 6}, \href{http://arxiv.org/abs/2201.07241}{{\tt arXiv:2201.07241 [astro-ph.CO]}}.

\bibitem{Mukherjee:2020vkx}
P.~Mukherjee and N.~Banerjee, ``{Revisiting a non-parametric reconstruction of the deceleration parameter from combined background and the growth rate data},'' \href{http://dx.doi.org/10.1016/j.dark.2022.100998}{{\em Phys. Dark Univ.} {\bf 36} (2022)  100998}, \href{http://arxiv.org/abs/2007.15941}{{\tt arXiv:2007.15941 [astro-ph.CO]}}.

\bibitem{Gomez-Valent:2018hwc}
A.~G\'omez-Valent and L.~Amendola, ``{$H_0$ from cosmic chronometers and Type Ia supernovae, with Gaussian Processes and the novel Weighted Polynomial Regression method},'' \href{http://dx.doi.org/10.1088/1475-7516/2018/04/051}{{\em JCAP} {\bf 04} (2018)  051}, \href{http://arxiv.org/abs/1802.01505}{{\tt arXiv:1802.01505 [astro-ph.CO]}}.

\bibitem{Haridasu:2018gqm}
B.~S. Haridasu, V.~V. Lukovi\'c, M.~Moresco, and N.~Vittorio, ``{An improved model-independent assessment of the late-time cosmic expansion},'' \href{http://dx.doi.org/10.1088/1475-7516/2018/10/015}{{\em JCAP} {\bf 10} (2018)  015}, \href{http://arxiv.org/abs/1805.03595}{{\tt arXiv:1805.03595 [astro-ph.CO]}}.

\bibitem{DESI:2024mwx}
{\bf DESI} Collaboration, A.~G. Adame {\em et al.}, ``{DESI 2024 VI: cosmological constraints from the measurements of baryon acoustic oscillations},'' \href{http://dx.doi.org/10.1088/1475-7516/2025/02/021}{{\em JCAP} {\bf 02} (2025)  021}, \href{http://arxiv.org/abs/2404.03002}{{\tt arXiv:2404.03002 [astro-ph.CO]}}.

\bibitem{DESI:2024lzq}
{\bf DESI} Collaboration, A.~G. Adame {\em et al.}, ``{DESI 2024 IV: Baryon Acoustic Oscillations from the Lyman alpha forest},'' \href{http://dx.doi.org/10.1088/1475-7516/2025/01/124}{{\em JCAP} {\bf 01} (2025)  124}, \href{http://arxiv.org/abs/2404.03001}{{\tt arXiv:2404.03001 [astro-ph.CO]}}.

\bibitem{DESI:2024uvr}
{\bf DESI} Collaboration, A.~G. Adame {\em et al.}, ``{DESI 2024 III: baryon acoustic oscillations from galaxies and quasars},'' \href{http://dx.doi.org/10.1088/1475-7516/2025/04/012}{{\em JCAP} {\bf 04} (2025)  012}, \href{http://arxiv.org/abs/2404.03000}{{\tt arXiv:2404.03000 [astro-ph.CO]}}.

\bibitem{DESI:2024aqx}
{\bf DESI} Collaboration, R.~Calderon {\em et al.}, ``{DESI 2024: reconstructing dark energy using crossing statistics with DESI DR1 BAO data},'' \href{http://dx.doi.org/10.1088/1475-7516/2024/10/048}{{\em JCAP} {\bf 10} (2024)  048}, \href{http://arxiv.org/abs/2405.04216}{{\tt arXiv:2405.04216 [astro-ph.CO]}}.

\bibitem{DESI:2016fyo}
{\bf DESI} Collaboration, A.~Aghamousa {\em et al.}, ``{The DESI Experiment Part I: Science,Targeting, and Survey Design},'' \href{http://arxiv.org/abs/1611.00036}{{\tt arXiv:1611.00036 [astro-ph.IM]}}.

\bibitem{DESI:2024kob}
{\bf DESI} Collaboration, K.~Lodha {\em et al.}, ``{DESI 2024: Constraints on physics-focused aspects of dark energy using DESI DR1 BAO data},'' \href{http://dx.doi.org/10.1103/PhysRevD.111.023532}{{\em Phys. Rev. D} {\bf 111} (2025) no.~2, 023532}, \href{http://arxiv.org/abs/2405.13588}{{\tt arXiv:2405.13588 [astro-ph.CO]}}.

\bibitem{Erices:2019mkd}
C.~Erices, E.~Papantonopoulos, and E.~N. Saridakis, ``{Cosmology in cubic and $f(P)$ gravity},'' \href{http://dx.doi.org/10.1103/PhysRevD.99.123527}{{\em Phys. Rev. D} {\bf 99} (2019) no.~12, 123527}, \href{http://arxiv.org/abs/1903.11128}{{\tt arXiv:1903.11128 [gr-qc]}}.

\bibitem{Asimakis:2022mbe}
P.~Asimakis, S.~Basilakos, and E.~N. Saridakis, ``{Building cubic gravity with healthy and viable scalar and tensor perturbations},'' \href{http://dx.doi.org/10.1140/epjc/s10052-024-12554-7}{{\em Eur. Phys. J. C} {\bf 84} (2024) no.~2, 207}, \href{http://arxiv.org/abs/2212.12494}{{\tt arXiv:2212.12494 [gr-qc]}}.

\bibitem{Kofinas:2014owa}
G.~Kofinas and E.~N. Saridakis, ``{Teleparallel equivalent of Gauss-Bonnet gravity and its modifications},'' \href{http://dx.doi.org/10.1103/PhysRevD.90.084044}{{\em Phys. Rev.} {\bf D90} (2014)  084044},
\href{http://arxiv.org/abs/1404.2249}{{\tt arXiv:1404.2249 [gr-qc]}}.

\bibitem{Weinberg:2008zzc}
S.~Weinberg, {\em {Cosmology}}.
\newblock 2008.

\bibitem{Eisenstein:2005su}
{\bf SDSS} Collaboration, D.~J. Eisenstein {\em et al.}, ``{Detection of the Baryon Acoustic Peak in the Large-Scale Correlation Function of SDSS Luminous Red Galaxies},'' \href{http://dx.doi.org/10.1086/466512}{{\em Astrophys. J.} {\bf 633} (2005)  560--574},
\href{http://arxiv.org/abs/astro-ph/0501171}{{\tt arXiv:astro-ph/0501171 [astro-ph]}}.

\bibitem{2dFGRS:2005yhx}
{\bf 2dFGRS} Collaboration, S.~Cole {\em et al.}, ``{The 2dF Galaxy Redshift Survey: Power-spectrum analysis of the final dataset and cosmological implications},'' \href{http://dx.doi.org/10.1111/j.1365-2966.2005.09318.x}{{\em Mon. Not. Roy. Astron. Soc.} {\bf 362} (2005)  505--534}, \href{http://arxiv.org/abs/astro-ph/0501174}{{\tt arXiv:astro-ph/0501174}}.

\bibitem{Alcock:1979mp}
C.~Alcock and B.~Paczynski, ``{An evolution free test for non-zero cosmological constant},'' \href{http://dx.doi.org/10.1038/281358a0}{{\em Nature} {\bf 281} (1979)  358--359}.

\bibitem{Seo:2003pu}
H.-J. Seo and D.~J. Eisenstein, ``{Probing dark energy with baryonic acoustic oscillations from future large galaxy redshift surveys},'' \href{http://dx.doi.org/10.1086/379122}{{\em Astrophys. J.} {\bf 598} (2003)  720--740}, \href{http://arxiv.org/abs/astro-ph/0307460}{{\tt arXiv:astro-ph/0307460}}.

\bibitem{Bahamonde:2024efl}
S.~Bahamonde and J.~Gigante~Valcarcel, ``{Stability in cubic metric-affine gravity},'' \href{http://dx.doi.org/10.1103/PhysRevD.111.084058}{{\em Phys. Rev. D} {\bf 111} (2025) no.~8, 084058}, \href{http://arxiv.org/abs/2411.12954}{{\tt arXiv:2411.12954 [gr-qc]}}.

\bibitem{Aoki:2023sum}
K.~Aoki, S.~Bahamonde, J.~Gigante~Valcarcel, and M.~A. Gorji, ``{Cosmological perturbation theory in metric-affine gravity},'' \href{http://dx.doi.org/10.1103/PhysRevD.110.024017}{{\em Phys. Rev. D} {\bf 110} (2024) no.~2, 024017}, \href{http://arxiv.org/abs/2310.16007}{{\tt arXiv:2310.16007 [gr-qc]}}.

\bibitem{Yo:2001sy}
H.-J. Yo and J.~M. Nester, ``{Hamiltonian analysis of Poincar\'e gauge theory: Higher spin modes},'' \href{http://dx.doi.org/10.1142/S0218271802001998}{{\em Int. J. Mod. Phys. D} {\bf 11} (2002)  747--780}, \href{http://arxiv.org/abs/gr-qc/0112030}{{\tt arXiv:gr-qc/0112030}}.

\bibitem{Yo:1999ex}
H.-j. Yo and J.~M. Nester, ``{Hamiltonian analysis of Poincare gauge theory scalar modes},'' \href{http://dx.doi.org/10.1142/S021827189900033X}{{\em Int. J. Mod. Phys. D} {\bf 8} (1999)  459--479}, \href{http://arxiv.org/abs/gr-qc/9902032}{{\tt arXiv:gr-qc/9902032}}.

\bibitem{Shie:2008ms}
K.-F. Shie, J.~M. Nester, and H.-J. Yo, ``{Torsion Cosmology and the Accelerating Universe},'' \href{http://dx.doi.org/10.1103/PhysRevD.78.023522}{{\em Phys. Rev.} {\bf D78} (2008)  023522},
\href{http://arxiv.org/abs/0805.3834}{{\tt arXiv:0805.3834 [gr-qc]}}.

\end{thebibliography}\endgroup

\end{document}